\documentclass[12pt,preprint]{aastex}

\newcommand{\be}{\begin{equation}}
\newcommand{\ee}{\end{equation}}
\newcommand{\bea}{\begin{eqnarray}}
\newcommand{\eea}{\end{eqnarray}}
\newcommand{\al}{\alpha}
\newcommand{\bt}{\beta}
\newcommand{\gm}{\gamma}
\newcommand{\Gm}{\Gamma}
\newcommand{\dl}{\delta}
\newcommand{\Dl}{\Delta}
\newcommand{\eps}{\epsilon}

\newcommand{\kp}{\kappa}
\newcommand{\lm}{\lambda}

\newcommand{\rh}{\rho}
\newcommand{\sg}{\sigma}

\newcommand{\om}{\omega}
\newcommand{\Om}{\Omega}

\newcommand{\rarrow}{\rightarrow}
\newcommand{\nn}{\nonumber}

\shorttitle{mhd vs gwaves}
\shortauthors{Kleidis, Kuiroukidis, and Papadopoulos }

\begin{document}

\title{Gravitational versus magnetohydrodynamic waves in curved spacetime, in the presence of large-scale magnetic fields}

\author{Kostas Kleidis$^{1}$, Apostolos Kuiroukidis$^{1}$ \\ and Demetrios B. Papadopoulos$^{2}$}
\affil{$^{1}$Department of Mechanical Engineering, International Hellenic University \\ (Serres Campus), 62124 Serres, Greece \\ $^{2}$ Department of Physics, Section of Astrophysics, Astronomy \& Mechanics,\\ Aristotle University of Thessaloniki, 54124 Thessaloniki, Greece}

\altaffiltext{}{E-mails: kleidis@ihu.gr, apostk@ihu.gr, papadop@astro.auth.gr}

\begin{abstract}
The general-relativistic (GR) magnetohydrodynamic (MHD) equations for a conductive plasma fluid are derived and discussed in the curved spacetime described by Thorne's metric tensor, i.e., a family of cosmological models with inherent anisotropy due to the existence of an ambient, large-scale magnetic field. In this framework, it is examined whether the magnetized plasma fluid that drives the evolution of such a model can be subsequently excited by a transient, plane-polarized gravitational wave (GW) or not. To do so, we consider the associated set of the perturbed equations of motion and integrate it numerically, to study the evolution of instabilities triggered by the GW propagation. In particular, we examine to what extend perturbations of the electric- and/or the magnetic field can be amplified due to a potential energy transfer from the GW to the electromagnetic (EM) degrees of freedom. The evolution of the perturbed quantities depends on four free parameters, namely, the conductivity of the fluid, $\sg$, the speed-of-sound-square, $\frac{1}{3}< \left ( \frac{C_s}{c} \right )^2 \equiv \gm < 1$, which in this model may serve also as a measure of the inherent anisotropy, the GW frequency, $\om_{g}$, and the associated angle of propagation with respect to the direction of the magnetic field, $\theta$. We find that, the GW propagation in the anisotropic magnetized medium under consideration does excite several MHD modes, in other words, there is energy transfer from the gravitational to the EM degrees of freedom that could result in the acceleration of charged particles at the spot and in the subsequent damping of the GW.
\end{abstract}


\section{Introduction}

A long time ago (i.e., back at the dinosaurs era), in a galaxy far-far away (NGC 4993) two neutron stars merged, emitting {\em gamma} rays and {\em gravitational waves} (GWs). On August 17, 2017, these two messengers reached the Earth (Abbott et al. 2017a). The GWs were detected (GW 170817) by the advanced LIGO and VIRGO interferometers (Abbott et al. 2017b, c) and the associated {\em gamma} rays (GRB 170817A) by the {\em Fermi} space telescope (Goldstein et al. 2017, Savchenko et al. 2017). The GW 170817 event and (most important) its verification by the associated {\em gamma} ray burst (GRB), not only has resulted in one more validation of General Relativity (GR), but, in fact, has revealed a powerful probe of the Universe exploration, the GW Astronomy. 

Apart from wobbling and merging neutron stars, other potential sources of GWs are coalescing compact binaries or black holes, quasi-normal modes of ringing black holes, the spinning-down effect observed in {\em magnetars} due to their enormous magnetic field, etc. (see, e.g., Duncan 1998, Maggiorre 2000, Anderson and Kokkotas 2001, Duez et al. 2001, Shibata and Urgu 2001). In each and everyone of those cases, a potential interaction between GWs and the astrophysical plasma being present at the spot not only could play an essential role to the final outcome of the messenger's profile, but it could also give rise to new phenomena of particular interest (see, e.g., Dermer et al. 1996, Hoshino et al. 1992, Papadopoulos et al. 2001), as, e.g., the excitation of MHD modes - especially of the magnetosonic waves (MSWs) - by the GW, i.e., the conversion of gravitational energy into EM energy (and vice versa). To search for potential resonances in the interaction between MHD modes of magnetized plasma and GWs in a curved spacetime background would be an essential first step towards that direction. An indicative, though incomplete literature on this subject includes derivation of the exact GR equations for finite amplitude MHD waves (Papadopoulos and Esposito 1982), GWs versus MSWs in the covariant formalism (Moortgat and Kuijpers 2003) and in the $3+1$ orthonormal tetrad description (Moortgat and Kuijpers 2004), coupling of GWs with EM waves in magnetized vacuum (Gertsenshtein 1961, Lupanov 1967, Boccaletti et al. 1970, Zel'dovich 1973, Gerlach 1974), propagation of GWs in magnetized plasma (Macedo and Nelson 1983, Kleidis et al. 1993, 1995, Anastasiadis et al. 1997), coupling of GWs with magnetized plasma in a Friedmann - Robertson - Walker (FRW) Universe (Marklund et al. 2000a, b), non-linear GW interaction with plasma (Brodin et al. 2000), frequency conversion by GWs (Brodin et al. 2001), exact spherically-symmetric MHD solutions in GR (Carot and Tupper 1999), evolution of gravitational instabilities in magnetized plasma (Hacyan 1983, Kleidis et al. 2008), dynamo effects in magnetized cosmologies (Kleidis et al. 2007), MHD perturbations in Bianchi Type I models (Fennely 1980, Jedamzik et al. 1998, 2000), large-scale magnetic fields and MHD phenomena in the early Universe (Thorne 1967, Jacobs 1968, 1969), MHD processes in the vicinity of central galactic engines (Dimmelmeier et al. 2002, Fryer et al. 2002, Baumgarte and Shapiro 2003), parametric resonant acceleration of particles (Kleidis et al. 1996) and/or amplification of EM plasma waves in a dispersive GW background (Mahajan and Asenjo 2022), and so on.

In spite of the wealth of references, a comprehensive study regarding the excitation of MHD modes (and their subsequent temporal evolution) by GWs is far from been exhausted. The various approaches considered so far, involve the interaction between gravitational and MHD waves mainly in two cases, i.e., either in flat spacetime or in an almost maximally-symmetric FRW cosmological model. In other words, it is admitted that the external magnetic field is too weak to destroy the spacetime homogeneity and isotropy. Yet, there exist several cases, either of astrophysical interest or in the primordial stages of the Universe evolution, where strong, ambient magnetic fields can have an important effect on the local spacetime structure (see, e.g., Moortgat and Kuijpers 2003, and references therein). In fact, as long as the magnetic field coherence length is larger or comparable to the causality horizon, isotropy is lost and an anisotropic background (where plasma and GWs co-exist and interact) must be taken into account to guarantee proper treatment (Tsagas and Maartens 2000). In fact, there is a class of cosmological models in which the magnetic field is inherently encapsulated to the spacetime geometry and results in its anisotropic evolution, the so-called Thorne's model (Thorne 1967, Misner et al. 1973).

In this article, we study the interaction between gravitational and MHD waves in a conductive (resistive) magnetized medium that drives the evolution of the anisotropic cosmological model described by Thorne's metric tensor. In Section 2, we set up the GR framework regarding the propagation of GWs in an anisotropic curved spacetime in the presence of an ambient EM filed. In Section 3, we derive the complete, self-consistent set of equations of motion for the perurbed MHD and GW modes. By virtue of a particular TT-gauge-like condition, only three non-zero components of the metric perturbations remain relevant. In Section 4, we perform a numerical study of the associated system of equations, using a fifth-order Runge - Kutta - Fehlberg temporal integration scheme of variable step. The corresponding results suggest that, the propagation of GWs in the anisotropic magnetized plasma so considered does excite several MHD modes, even if, initially, all the perturbed quantities were {\em null}, i.e., their temporal initial conditions had been set equal to zero. In other words, energy transfer from the gravitational to the EM degrees of freedom does take place, depending on the conductivity of the plasma fluid (resistive instabilities), the anisotropy of the curved spacetime (anisotropic instabilities), the frequency of the GW (dispersive instabilities) and the associated angle of propagation with respect to the direction of the ambient magnetic field (resonant instabilities). Finally, we conclude in Section 5. In what follows, we use the system of units where the velocity of light, $c$, Boltzmann's constant, $k_B$, and Newton's gravitational constant, $G$, are all equal to unity, i.e., $c = k_B = G = 1$.

\section{Propagation of GWs in magnetized anisotropic cosmologies}

We consider an axisymmetric Bianchi Type I cosmological model, the line-element of which is written in the form \be ds^2 = - dt^2 + A^2 (t) \left [ dx^2 + dy^2 \right ] + W^2 (t) dz^2 , \ee where $A(t)$ and $W(t)$ are the dimensionless scale factors. The anisotropy along the $\hat{z}$-direction is due to an ambient magnetic field of the form $\vec{H} = H(t) \hat{z}$, where $H(t) \equiv F_{\; \; y}^x$, with $F_{\; \; y}^x$ being the {\em EM field tensor}, and $\hat{z} = W^{-1} \partial_z$. Assuming that the curved background given by Eq. (1) is filled with a perfect (though conductive) fluid with equation of state $p = \gm \rh$, where $\rh$ is the {\em rest-mass density}, $p$ is the {\em pressure}, and $\frac{1}{3} < \gm < 1$ denotes the {\em speed of sound square}, $C_s^2$, the associated Einstein-Maxwell equations yield (Thorne 1967) \bea && A(t) = t^{1/2}, \; \; W(t) = t^{\ell}, \; \; \; \ell \equiv \frac{(1 - \gm)}{(1 + \gm)} \: , \\ && H(t) = \frac{(1 - \gm)^{1/2}(3 \gm - 1)^{1/2}}{2 (1 + \gm) } \times \frac{1}{t} \: , \\ &&  \rh (t) = \frac{(3 - \gm )}{16 \pi (1 + \gm)^{2}} \times \frac{1}{ t^{2}} \: . \eea However, since in the original derivation of Thorne's model the ambient magnetic field corresponds to the mixed component of the EM field tensor, in what follows the background magnetic field reads \be H_{(B)} \equiv g^{yy} H(t) = \frac{H(t)}{t} = \frac{(1 - \gm)^{1/2}(3 \gm - 1)^{1/2}}{2 (1 + \gm) } \times \frac{1}{t^2} \: . \ee Notice that, for $\gm \rarrow \frac{1}{3}$, the {\em magnetic field strength}, $H(t)$, along with the associated anisotropy vanish. In fact, for $\gm \rarrow \frac{1}{3}$, $\ell \rarrow \frac{1}{2}$ and we obtain the isotropic, FRW radiation-dominated Universe. In this context, Thorne's model can be considered as an extension to the Standard Model of the Universe expansion history and, in particular, as an alternative to the radiation era of the early Universe in the case where large-scale magnetic fields might have ever played an important role in cosmic expansion. Actually, we do not know. 

On the other hand, in the {\em stiff matter} approach (i.e., for $\gm \rarrow 1$) the inherent anisotropy of Eq. (1) not ony remains active, but it becomes even more prominent, since, although $H \rarrow 0$, the $\hat{z}$-axis becomes static $(W \rarrow 1)$, resulting in a {\em pancake} model. Hence, the speed-of-sound-square parameter, $\gm$, may serve also as a measure of the inherent anisotropy. The exact solution given by Eqs. (1) - (4) determines the class of Thorne's anisotropic magnetized cosmologies (Thorne 1967), which will serve as our background metric, $g_{\mu \nu }^{(B)}$, i.e., in what follows, tensor indices are raised and lowered using this metric tensor. In Thorne's model, the {\em Alfv\'{e}n group velocity} of the MHD waves propagating in the interior of the conductive cosmic fluid along the $\hat{z}-$ axis remains constant, namely, \be u_A^2 \equiv \frac{H^2}{4 \pi \rh} = \frac{(1 - \gm) (3 \gm - 1)}{(3 - \gm)} \: . \ee 
 
Let us consider a plane-polarized GW propagating  in the aforementioned model, at an angle $\theta$, with respect to the direction of the magnetic field. In this case, the associated spacetime metric reads \be g_{\mu \nu } = g_{\mu \nu }^{(B)} + h_{\mu \nu} \: , \ee where $\vert h_{\mu \nu} \vert \ll 1$ and Greek indices refer to the four-dimensional spacetime (in accordance, Latin indices refer to the three-dimensional spatial section). Following Misner et al. (1973), we admit that $\dl g_{\mu \nu } = h_{\mu \nu}$, hence, $h^{\al \bt} \equiv g^{\al \mu}_{(B)} \: g^{\bt \nu}_{(B)} \: h_{\mu \nu }$, and $\dl g^{\mu \nu} = - h^{\mu \nu}  = - g^{\mu \al}_{(B)} \: g^{\nu \bt}_{(B)} \: h_{\al \bt}$, where repeated upper and lower indices denote summation. Without loss of generality, we may restrict ourselves to a GW propagating on the $(y,z)$-plane. In this case, GW propagation takes place along the $\xi$-direction, which, together with its normal one, the $v$-direction, are determined by \bea \xi & = & z \: \sin \theta + y \: \cos \theta \: , \nn \\ v & = & z \: \cos \theta - y \: \sin \theta \: . \eea In view of Eqs. (8), the associated covariant components of the metric tensor (1) are given by \bea g_{\xi \xi } & = & \cos^{2} \theta A^{2} + \sin^{2} \theta W^{2} \: , \nn \\ g_{vv} & = & \cos^{2} \theta W^{2} + \sin^{2} \theta A^{2} \: , \nn \\ g_{\xi v} & = & \sin \theta \cos \theta (W^{2}-A^{2}) \eea and their contravariant counterparts are written in the form \bea g^{\xi \xi } & = & \left ( \frac{\sin^2 \theta }{W^{2}}+\frac{\cos^{2} \theta}{A^{2}}\right ) \: , \nn \\ g^{vv} & = & \left ( \frac{\sin^{2} \theta }{A^{2}} +\frac{\cos^{2} \theta}{W^{2}} \right ) \: , \nn \\ g^{\xi v} & = & \sin \theta \cos \theta \left ( \frac{1}{W^{2}} - \frac{1}{A^{2}} \right ) \: , \eea for which the following (auxiliary) conditions hold \bea g_{\xi \xi } g_{vv} - g_{\xi v}^{2} & = & A^{2} W^{2} \: , \\  g^{\xi \xi } g^{vv} - \left ( g^{\xi v} \right )^{2} & = & \frac{1}{A^{2} W^{2}} \: . \eea Consequently, from now on, all quantities depend on $(t, \xi)$ and the angle of propagation ranges from zero to $\frac{\pi}{2}$. For $\theta = 0$ , the GW propagates {\em normal} to the direction of the magnetic field, while, for $\theta = \frac{\pi}{2}$, {\em parallel} to that direction. 

The GW's equation of propagation in the background determined by Eqs. (1) - (4) is given by (see, e.g., Misner et al. 1973) \be (h_{\mu \nu})_{|\al }^{\al } + 2 {\cal R}^{(B)}_{\mu \al \nu \bt} h^{\al \bt } = g_{(B)}^{\al \bt} \left ( t_{\al \mu } h_{\nu \bt} + t_{\al \nu} h_{\mu \bt} \right ) \: , \ee where the vertical bar denotes {\em covariant derivative} with respect to the background metric, ${\cal R}^{(B)}_{\mu \al \nu \bt}$ is the associated {\em Riemann curvature tensor}, and $t_{\al \bt}$ is the {\em source tensor}, i.e., the part including all the non-gravitational fields, given by \be t_{\al \mu} \equiv 8 \pi \left ( T_{\al \mu } - \frac{1}{2} T g_{\al \mu}^{(B)} \right ) \: . \ee In Eq. (14), $T_{\al \mu}$ is the {\em stress-energy tensor} of model's (1) matter content and $T \equiv  g^{\mu \nu}_{(B)} T_{\mu \nu}$ denotes its trace. As regards the background metric itself, the Einstein field equations are written in the form \be {\cal R}_{\al \bt}^{(B)} = t_{\al \bt} \: , \ee with ${\cal R}_{\al \bt}^{(B)}$ being the associated {\em Ricci tensor}. In a cosmological model filled with perfect fluid in the presence of an ambient EM field, the stress-energy tensor is decomposed to \bea T^{\mu \nu} = T^{\mu \nu}_{(em)} + T^{\mu \nu}_{(fluid)} & = & \frac{1}{4\pi } \left ( F^{\mu \al} F^{\nu \bt} g_{\al \bt} - \frac{1}{4} g^{\mu \nu} F^2 \right ) + \nn \\ & + & \left [ (\rho + p) \: u^{\mu } u^{\nu }+ p\: g^{\mu \nu} \right ] \: , \eea where $u^{\al}_{(B)} = (1, 0, 0, 0)$ is the fluid's {\em four-velocity} at rest with respect to the comoving frame. In this case, the {\em Faraday tensor}, $F^{\mu \nu}$, of the associated EM field is given by \bea F^{\mu \nu} = \left [
\begin{tabular}{cccc}
0,&$E^x$,&$E^y$,&$E^z$ \\
$- E^x$,&0,&$B_z$,&$- B_y$\\
$- E^y$,&$- B_z$,&0,&$B_x$\\
$- E^z$,&$B_y$,&$- B_x$&0
\end{tabular} \right ] \: , \eea where $B_i$ $(i = x, y, z)$ is {\em magnetic induction}. Notice that, as long as there is no magnetization field, the quantities $H_i$ and $B_i$ coincide up to a constant. Accordingly, the components of the electric field are given by $E^j = F^{j \mu} u_{\mu }^{(B)}$ and their magnetic counterparts are written in the form $B_i = \frac{1}{2} \eps_{i j k m} F^{jk} u_{(B)}^m$. The stress-energy tensor given by Eq. (16) satisfies the {\em conservation law} \be T^{\mu \nu}_{|\nu } = 0 \ee and Maxwell's equations in curved spacetime read (see, e.g., Misner et al. 1973) \bea F^{\al \bt}_{|\bt } & = & 4 \pi J^{\al} \: , \nn \\ F_{\al \bt |\gm} + F_{\bt \gm|\al} + F_{\gm \al|\bt} & = & 0 \: , \\ J^{\al }_{|\al }& = & 0 \: , \nn \eea where $J^{\al}$ is {\em current density}. In this article, we admit that the cosmic fluid representing the matter content of Thorne's model corresponds to a locally neutral, two-component plasma, in which the local neutrality is achieved due to the mobility of the lighter ion-species. Accordingly, the current density can be obtained from the invariant form of Ohm's law, as \be J^{\al} = \rh_e u^{\al} + \sg F^{\al \bt} u_{\bt} \: , \ee with $\sg$ being the {\em conductivity} (see, e.g., Jackson, 1975). In Eq. (20), $\rh_e$ is the local {\em charge density}, the unperturbed value of which equals to zero, i.e., $\rh_e = 0$. However, provided that the conductivity of the fluid is {\em finite}, we may admit that, locally, $\dl \rh_e (t, \vec{r}) \neq 0$  (see, e.g.,  Dendy 1990). In fact, in what follows, we assume that the conductivity of the magnetized cosmic plasma fluid not only is finite, but it also remains {\em constant} in time. However, in general, the conductivity can vary with time, following the Spitzer relation (see, e.g., Krall and Trivielpiece 1973) \be \sigma =10^{2} \left ( \frac{T}{eV} \right )^{3/2} \; \; (sec)^{-1} , \ee where $T$ is the plasma temperature. According to Standard Model, the Universe evolution could be driven by plasma at the time era between the epoch of {\em matter-radiation equality}, that took place at $t_{eq} \simeq 10^{11} \; sec$, and the {\em recombination epoch} ($t_{rec} \simeq 10^{13} \; sec$), at which the temperature drops to the point where electrons and nuclei can form stable atoms. In this time interval, we have $T = 10^{5.4} \: t^{-1/2} \: eV$ (see, e.g., Kolb and Turner 1990). Consequently, \be \sg \approx \frac{10^{10}}{t^{3/4}} \; \; sec^{-1} \: , \ee hence, for $t_{eq} \leq t \leq t_{rec}$ the conductivity of the cosmic plasma fluid ranges from $\sg_{rec} \simeq 2 \; sec^{-1}$ to $\sg_{eq} \simeq 55 \; sec^{-1}$. For this reason, in what follows, $\sg$ is going to be treated as a constant parameter, that takes values in the range $1 \leq \sg \leq 50$ $sec^{-1}$. In view of the above setting, we may, now, proceed to derive the resistive MHD and GW equations of propagation in the background model given by Eqs. (1) - (4). 

Notice that, upon consideration of the condition $u^{\al} u_{\al} = - 1$, the perturbed four-velocity vector is written in the form \be u^{\al} (t, \xi) = (1, \dl u^x, \dl u^y, \dl u^z) \: . \ee 

In order to analytically express (and solve) Eqs. (13), we need to impose a {\em gauge condition} on $h_{\mu \nu}$. In the curved background of Thorne's model, such a condition would be \be (h_{\mu}^{\al })_{|\al } = 0 \; \; \; and \; \; \; h = 0, \ee where $h$ is the trace of $h^{\mu \nu}$. The gauge condition given by Eq. (24) is the closest to the {\em transverse-traceless}, (i.e., TT-gauge) condition we can impose in the curved background given by Eq. (1). The first of Eqs. (24) can be cast in the form \be \frac{1}{\sqrt{-g}} \left ( \sqrt{-g} \; h_{\mu }^{\al } \right)_{,\al } - \Gm _{\kp \mu \al} g^{\kp \lm } g^{\al \bt } h_{\lm \bt } = 0 \: , \ee where $g = - A^4 W^2 = - t^{2 (\ell +1)}$ is the determinant of the metric tensor (1), $\Gm_{\kp \mu \al}$ are the associated Christoffel symbols, and the comma denotes partial differentiation. Notice that, in Eqs. (25), all the components of the metric tensor involved correspond to those of the background metric. On the approach of three arbitrary functions, $F^{(i)} (t, \xi ) \; (i = 1, 2, 3)$, Eqs. (24) can be solved explicitly in terms of $h_{00}$ and $h_{xv}$, as follows \be h_{x0}=\frac{F^{(1)}_{,\xi }}{\sqrt{-g}}, \; \; \; h_{\xi 0}=\frac{F^{(2)}_{,\xi }}{\sqrt{-g}}, \; \; \; h_{v0}=\frac{F^{(2)}_{,\xi }}{\sqrt{-g}} \: , \ee
\be h_{x \xi}=\frac{F^{(1)}_{,0}}{\sqrt{-g} \: g^{\xi \xi }} - \frac{g^{\xi v}}{g^{\xi \xi }}h_{xv} \: , \ee 
\be h_{\xi \xi }=\frac{F^{(2)}_{,0}}{\sqrt{-g} \: g^{\xi \xi }} -\frac{g^{\xi v}}{g^{\xi \xi }}h_{\xi v},\; \; \; \; h_{vv}=\frac{F^{(3)}_{,0}}{\sqrt{-g} \: g^{\xi v}} -\frac{g^{\xi \xi}}{g^{\xi v}}h_{\xi v} \: , \ee and
\be g^{xx}h_{xx}=h_{00}-\frac{F^{(2)}_{,0}}{\sqrt{-g}}- \frac{F^{(3)}_{,0}} {\sqrt{-g}}\frac{g^{vv}}{g^{\xi v}} + \frac{h_{\xi v}}{A^{2}W^{2}g^{\xi v}} \: , \ee  where $h_{\xi v}$ is also related to $h_{00}$, through the condition \bea && \frac{\cos^{2} \theta}{A^{4}W^{2} g^{\xi \xi } g^{\xi v}} \left ( \frac{\dot{A}}{A} - \frac{\dot{W}}{W} \right ) h_{\xi v} = -\frac{(\sqrt{-g} \: h_{00})_{,0}}{\sqrt{-g}} - \left ( \frac{\dot{A}}{A} \right ) h_{00}+ \frac{g^{\xi \xi }F^{(2)}_{,\xi \xi }+g^{\xi v}F^{(3)}_{,\xi \xi }} {\sqrt{-g}} \nn \\ && + \frac{F^{(2)}_{,0}}{\sqrt{-g} \: g^{\xi \xi }}\frac{\sin^2 \theta }{W^{2}} \left ( \frac{\dot{A}}{A}-\frac{\dot{W}}{W} \right )+ \frac{g^{\xi \xi }F^{(2)}_{,\xi \xi }+g^{\xi v}F^{(3)}_{,\xi \xi }} {\sqrt{-g}}+ \frac{F^{(3)}_{,0}}{\sqrt{- g} \: g^{\xi v}}\frac{\cos^2 \theta}{W^{2}} \left ( \frac{\dot{A}}{A}-\frac{\dot{W}}{W} \right ) \eea and the dot denotes differentiation with respect to cosmic time, $t$. 

In view of Eqs. (26) - (30), Eqs. (13) are left with only three unspecified (i.e., that cannot be set equal to zero independently) GW components, namely, $h_{x0}, \; h_{x \xi }$, and $h_{xv}$, accompanied by the {\em phase-space constraint} \be \frac{(A^{2}W h_{0x})_{,0}}{A^{2} W} =  \left ( \frac{\sin^{2} \theta}{W^{2}} + \frac{\cos^{2} \theta }{A^{2}} \right ) h_{x\xi ,\xi } + \sin \theta \cos \theta \left ( \frac{1}{W^{2}}-\frac{1}{A^{2}} \right ) h_{xv,\xi } \: , \ee a direct consequence of Eqs. (26) - (28). Along with Eqs. (13), we may now perturb also Eqs. (18) and (19), with respect to the metric, the EM field, and the cosmic fluid variables, to obtain the following closed, self-consistent set of perturbations' equations of propagation.

\section{The perturbations' equations of propagation in Thorne's model}
 
For clarity reasons, the perturbations' equations of propagation are sorted as follows:

\subsection{GW propagation in Thorne's model}

Along with the {\em phase-space constraint} given by Eq. (31), the propagation of the non-zero GW components in Thorne's model (1) is governed by the following set of equations \be \Box (h_{0x}) -2 \frac{\ddot{A}}{A}(h_{0x}) = -16\pi \gm \rho (h_{0x}) \: , \ee
\bea \Box (h_{x\xi }) & - & 2 \left ( \frac{\dot{A}}{A} \right ) \left [ \cos^{2}\theta \frac{\dot{A}}{A} + \sin^{2} \theta \frac{\dot{W}}{W} \right ] (h_{x\xi }) - 2 \left ( \frac{\dot{A}}{A} \right ) \sin \theta \cos \theta \left ( \frac{\dot{W}}{W} - \frac{\dot{A}}{A} \right ) (h_{xv}) \nn \\ & = & 8 \pi \left [ \rho (1-\gm ) + \cos^{2} \theta A^{4} \frac{H_{(B)}^{2}}{4\pi } \right ] (h_{x\xi }) - 8 \pi \sin \theta \cos \theta \left [ A^{4} \frac{H_{(B)}^{2}}{4\pi} \right ] (h_{xv}) , \eea and
\bea \Box (h_{xv}) & - & 2 \left ( \frac{\dot{A}}{A} \right ) \left [ \sin^{2} \theta \frac{\dot{A}}{A} + \cos^{2} \theta \frac{\dot{W}}{W} \right ] (h_{xv}) - 2 \left ( \frac{\dot{A}}{A} \right ) \sin \theta \cos \theta \left ( \frac{\dot{W}}{W} - \frac{\dot{A}}{A} \right ) (h_{x \xi }) \nn \\ & = & 8 \pi \left [ \rho (1-\gm ) + \sin^{2} \theta A^{4} \frac{H_{(B)}^{2}}{4 \pi} \right ] (h_{xv}) -  8 \pi \sin \theta \cos \theta \left [ A^{4} \frac{H_{(B)}^{2}}{4\pi} \right ] (h_{x\xi }) , \eea
where \be \Box f = - \partial_t^2 f + \left ( \frac{\sin^2 \theta} {W^2} + \frac{\cos^2 \theta}{A^2} \right ) \partial_{\xi }^2 f - \left ( \frac{2 \dot{A}}{A} + \frac{\dot{W}}{W} \right ) \partial_{t} f \: , \ee is the D' Alembert wave operator in the curved spacetime given by Eq. (1) and $f_{,0} = \partial_t f = \dot{f}$.

\subsection{The perturbed Maxwell equations}

Next, we consider the perturbed Maxwell equations, that give rise to the EM-field perturbations' equations of propagation. Perturbation of the first of Eqs. (19) with respect to metric and the EM variables, yields the following set of equations \be \cos \theta (\delta E^{y})_{,\xi } + \sin \theta (\delta E^{z})_{,\xi } = 4 \pi (\delta \rho_{e}) \: , \ee
\bea &-& (\delta E^{x})_{,0} + \cos \theta (\delta B_{z})_{,\xi } - \sin \theta (\delta B_{y})_{,\xi } - \nn \\ &-& (\delta E^{x}) \left ( \frac{2\dot{A}}{A}+\frac{\dot{W}}{W} \right ) = 4 \pi \sigma \left [ \delta E^{x}+A^{2}H_{(B)} \delta u^{y} \right ] \: , \eea
\bea &-&(\delta E^{y})_{,0} + \sin \theta (\delta B_{x})_{,\xi } - (\delta E^{y}) \left ( \frac{2 \dot{A}}{A} + \frac{\dot{W}}{W} \right ) = \nn \\ & = & 4 \pi \sigma \left [ \delta E^{y}-A^{2}H_{(B)} \delta u^{x} - H_{(B)}h_{x0} \right ] \: , \eea and
\be - (\delta E^{z})_{,0} - \cos \theta (\delta B_{x})_{,\xi } - (\delta E^{z}) \left ( \frac{2\dot{A}}{A}+\frac{\dot{W}}{W} \right ) = 4 \pi \sigma (\delta E^{z}) \: . \ee
Subsequently, perturbing the second of Eqs. (19), we obtain \be (A^{4} \delta B_{z})_{,0} - \cos \theta A^{2} (\delta E^{x})_{,\xi } = 0 \: , \ee
\be (A^{2}W^{2} \delta B_{y})_{,0} + \sin \theta A^{2} (\delta E^{x})_{,\xi }=0 \: , \ee
\bea (A^{2} W^{2} \delta B_{x})_{,0} & - & \sin \theta A^{2} (\delta E^{y})_{,\xi } + \cos \theta W^{2} (\delta E^{z})_{,\xi }+ \nn \\ & + & \sin \theta A^{2}H_{(B)}(h_{0x})_{,\xi } - \sin \theta [A^{2}H_{(B)}h_{x\xi }]_{,0} - \nn \\ & - & \cos \theta [A^{2}H_{(B)}h_{xv}]_{,0} = 0 \: , \eea and
\be \sin \theta A^{2}(\delta B_{z})_{,\xi } + \cos \theta W^{2} (\delta B_{y})_{,\xi } = 0 \: . \ee
Notice that, in view of Eq. (43), the choice \be \dl B_y = - \left ( \frac{\sin \theta}{\cos \theta} \right ) t^{(1-2 \ell)} \dl B_z \ee would make Eqs.(40) and (41) identical. In this case, the only remaining equations of this subset are Eqs. (40) and (42). The former can be cast in the more convenient  form \be (\dl B_z)_{, 0} + \frac{2}{t} (\dl B_z) - \frac{\cos \theta}{t} (\dl E^x)_{, \xi } = 0 \: . \ee Finally, perturbation of the last of Eqs. (19) with respect to the metric and the EM variables, results in \bea (\delta \rho _{e})_{,0} & + & \cos \theta \left [ \sigma \delta E^{y} - \sigma H_{(B)}h_{x0} - \sigma A^{2}H_{(B)}\delta u^{x} \right ]_{,\xi } + \sin \theta \left [ \sigma \delta E^{z} \right ]_{,\xi }+ \nn \\ & + & \left ( 2\frac{\dot{A}}{A}+\frac{\dot{W}}{W} \right ) \delta \rho_{e} = 0 \: . \eea

\subsection{The perturbed Euler equations}

To conclude with the perturbations' equations of propagation, we now perturb the conservation law given by Eq. (18), also with respect to the metric and the EM-field variables, to obtain the perturbation equations for the cosmic fluid variables, as follows \bea \partial_{t}(\delta \rho )& + &\frac{1}{4\pi } \left [ A^{4}H_{(B)}\delta B_{z} \right ]_{,0}+(1+\gm )\rho \left [ \cos \theta (\delta u^{y})_{,\xi } + \sin \theta (\delta u^{z})_{,\xi } \right ] - \nn \\ & - & \frac{1}{4\pi} \cos \theta A^{2}H_{(B)} (\delta E^{x})_{,\xi } + (1 + \gm ) \left ( \frac{2\dot{A}}{A}+\frac{\dot{W}}{W} \right ) \delta \rho + \nn \\ & + & 4 \left ( \frac{\dot{A}}{A} \right ) \left ( \frac{1}{4 \pi} A^{4} H_{(B)} \right ) \delta B_{z} = 0 \: , \eea
\bea (1+\gm )\partial_{t}(\rho \delta u^{x}) & + & \frac{1}{4\pi }[A^{2}H_{(B)}\delta E^{y}]_{,0}- \frac{1}{4\pi} \sin \theta A^{2}H_{(B)} (\delta B_{x})_{,\xi } + \gm \left [ \frac{\rho }{A^{2}}h_{0x} \right ]_{,0} - \nn \\ & - & \frac{1}{8\pi }[A^{2} H_{(B)}^2 h_{0x}]_{,0} + \left [ \frac{\rho }{A^{2}} + A^{2}\frac{H_{(B)}^{2}}{8\pi } \right ] (h_{0x})_{,0} + \nn \\ & + & \left ( 4\frac{\dot{A}}{A}+\frac{\dot{W}}{W} \right ) \left [ (1+\gm ) \rho \delta u^{x}+ \frac{1}{4\pi }A^{2}  H_{(B)}\delta E^{y} \right ] + \nn \\
& + & 2 \left ( \frac{\dot{A}}{A}\right) \gm \rho \frac{h_{0x}}{A^{2}} - 6 \left ( \frac{\dot{A}}{A} \right ) A^{2}\frac{H_{(B)}^{2}}{8\pi }h_{0x} = 0 \: , \eea
\bea (1+\gm) \partial _{t} (\rho \delta u^{y}) & - & \frac{1}{4\pi} [A^{2}H_{(B)}\delta E^{x}]_{,0} + \gm \frac{\cos \theta}{A^{2}}(\delta \rho )_{,\xi } + \nn \\ & + & \frac{1}{4\pi}A^{2}H_{(B)} [\cos \theta (\delta B_{z})_{,\xi } - \sin \theta (\delta B_{y})_{,\xi }] + \nn \\ & + & \left ( 4 \frac{\dot{A}}{A}+\frac{\dot{W}}{W} \right ) \left [ (1+\gm )\rho \delta u^{y}-\frac{1}{4\pi }A^{2} H_{(B)}\delta E^{x} \right ] = 0 \: , \eea and
\bea (1+\gm) \partial_{t} (\rho \delta u^{z}) & + & \gm \frac{\sin \theta }{W^{2}}(\delta \rho )_{,\xi } +\frac{1}{4 \pi} A^{2}H_{(B)} [- \cos \theta (\delta B_{y})_{,\xi } - \sin \theta \frac{A^{2}}{W^{2}}(\delta B_{z})_{,\xi }] + \nn \\ & + & \left ( 2\frac{\dot{A}}{A}+ 3 \frac{\dot{W}}{W} \right ) \left [ (1+\gm )\rho \delta u^{z} \right ] = 0 \: . \eea
Eqs. (32) - (50) constitute the complete, closed set of 14 perturbation equations that govern the evolution of the metric, the EM field, and the fluid variables involved, namely, $h_{x0}$, $h_{x \xi }$, and $h_{x v}$ [Eqs. (32) - (34)], $\dl \rh_e$, $\dl E^x$, $\dl E^y$, $\dl E^z$, $\dl B_x$, $\dl B_y$, and $\dl B_z$ [Eqs. (37) - (42) and (46)], and $\dl \rh$, $\dl u^x$, $\dl u^y$, and $\dl u^z$ [Eqs. (47) - (50)], respectively, along with three first-order (phase space) constraints, namely, Eqs. (31), (36), and (43). Although over-determined, this is a very complicated system of partial differential equations of inter-connected variables, to which several different classes of solutions may exist. Clearly, it is very difficult to deal with it analytically; hence, we focus on its numerical study. To do so, we need to point out that:

\begin{itemize}

\item As we have already stated, the lower temporal limit $(t_0)$ of numerical integration is taken to be the epoch of matter-radiation equality, which occurs at $t_0 \sim 10^{11} \; sec$ (see, e.g., Kolb and Turner 1990). Accordingly, we normalize time in units of $t_0$. For $t \geq t_0$, the plasma-dominated Universe goes on expanding and cooling until $t_{rec} \sim 10^{13} \; sec$, when the temperature drops to the point where electrons and nuclei can form stable atoms (recombination) and no plasma is left in the Universe, at all. Hence, the latest time at which plasma could play a role of cosmological significance is $t_{rec} = 100 \: t_0$, which will serve as the upper limit of numerical integration.Therefore, the limits of numerical integration are $1 \leq \frac{t}{t_0} \leq 100$.

\item To study resistive and/or other types of instability that might be triggered by the oblique propagation of a plane polarized GW within the anisotropic magnetized plasma fluid under consideration, we assume that all the EM-field and fluid perturbations involved correspond to {\em plane-wave-like forms}, \be \dl f (t, \xi) = \dl f (t) e^{\imath (k \xi - \int^t \om dt)} \: , \ee following the so-called {\em adiabatic approximation} (Zel’dovich 1979, Birrell and Davies 1982, Padmanabhan 1993), where $k$ is the {\em comoving wave-number}. In this context, the (slowly varying) time-dependent frequency of the various wave-forms, $\om (k, t)$, is defined by the {\em eikonal} $\Om = \int^t \om dt$ through the relation $\om = \frac{d \Om}{dt}$. To allow for potential {\em resonances} between the GW and the EM degrees of freedom, in what follows, we assume that $\om \approx \om_g$, so that $\om$ can be Taylor-expanded around $\om_g$. In this case, \be \om = \om_g + \frac{d \om}{dt} \left \vert_{\om_g} \right . t + \frac{d^2 \om}{dt^2} \left \vert_{\om_g} \right . t^2 + ... \ee and the eikonal $\Om$ results in \be \Om = \int_{t_0}^t \om dt =  \om_g \left ( \frac{t}{t_0} \right ) \left [ 1 + \frac{1}{200} \left ( \frac{t}{t_0} \right ) + \frac{1}{60000} \om_g \left ( \frac{t}{t_0} \right )^2 + ... \right ] \: , \ee i.e., during the whole time-interval of numerical integration it is slightly modulated around the quantity $\om_g \left ( \frac{t}{t_0} \right )$ (the main resonance). 

\item On the other hand, $\om$ as defined by Eqs. (51) and (52), has the usual meaning of the {\em angular frequency} of an oscillating process only in the short-wavelength (high-frequency) regime of the mode $k$ (see, e.g., Mukhanov et al. 1992). In other words, the wave description in curved spacetime makes sense only when the {\em physical wavelength} along the direction of propagation, $\lm_{ph} = \lm W(t)$, is smaller than the associated {\em horizon length}, $\ell_{H_W} = H_W^{-1}$, where $H_W = \frac{1}{W} \frac{dW}{dt}$ is the Hubble parameter along the anisotropic $\hat{z}-$direction. Accordingly, \be \lm_{ph} \leq \ell_{H_W} \Rightarrow \frac{k}{W(t)} \geq H_W \Rightarrow k \geq \frac{1 - \gm}{1 + \gm} \left ( \frac{t_0}{t} \right )^{\frac{2 \gm}{1 + \gm}} \: . \ee For  $1 \leq \frac{t}{t_0} \leq 100$, it suffices that $k \geq \frac{1 - \gm}{1 + \gm}$, measured in units of $c^{-1}$, i.e., $ \times 10^{-10} \: (cm)^{-1}$.

\item Eventually, at $t = t_0$, with the exception of GWs, all other perturbations are taken to admit $\dl f(t_0) = 0$. As regards metric perturbations, they correspond to pre-recombination cosmological GWs, the initial amplitude of which is given by \be \dl f_g (t_0) \equiv A = \al \sqrt {\frac{L}{2 \om_g}} \times 10^{-11} \; \; (cm) \: , \ee where $L$ is a characteristic length scale (see, e.g., Maggiore 2000). In the absence of any interaction, these GWs would manifest an amplitude of the order $A \sim 10^{-20}$ $cm$ at the present epoch. In Eq. (55), $\al = \{ 0.5, \; 0.9, \; 1.4, \; 1.9, \; 2.5 \}$ is the {\em normalized initial amplitude} of the GW. It is used to examine to which extend the GW temporal evolution might result in its amplitude reduction and, most important, what the corresponding effect on the EM potentials would be. We stress that a potential descending behaviour of the GW amplitude would signal an energy transfer from the gravitational to the EM degrees of freedom of the MHD system under consideration, leading to the excitation of the EM fields and the damping of the GW (see, e.g., Kleidis et al. 1995). In this case it is interesting to explore the role of the various free parameters involved. In fact, we shall examine the role of conductivity, $\sg$, and the inherent spacetime anisotropy, $\gm$, along with those of the GW frequency, $\om_g$, and the associated angle of propagation with respect to the direction of the background magnetic field, $\theta$. 

\end{itemize}

Accordingly, the 14 differential equations of propagation of the perturbed quantities, namely, Eqs. (32) - (34), (37) - (42), and (46) - (50) are integrated forward in time, using the fifth-order Runge - Kutta - Fehlberg computational scheme with variable integration step. In order to study the dependence of the evolution of the perturbation modes on a particular parameter, at each {\em numerical run} we keep all other parameters, i.e., exept the one under consideration, as constants. 

\section{Numerical evaluation of the perturbation modes}

We consider a plane-polarized GW of frequency $\om_g = 0.1$ $Hz$, propagating within a plasma fluid of conductivity $\sg = 10$ $sec^{-1}$, at an angle $\theta = 45^{\circ}$ with respect to the direction of an ambient magnetic field, $\vec{H} = H(t) \hat{z}$. The magnetized plasma so considered drives the evolution of the associated Thorne's model with $\gm = \left ( \frac{C_s}{c} \right )^2 = 0.6$. The temporal evolution of the GW component $h_{x0}$ (normalized over its initial amplitude) is presented in Fig. 1, for $\al = 1.4$ (the minus sign is nothing but a convention, denoting that $ \vert h_{\mu \nu} \vert $ decreases). In fact, the evolution of $h_{x0}$ is independent of $\al$, as a consequence of the constraint given by Eq. (31), something that is verified also by the numerical results. 

From Fig. 1, we observe that the metric perturbation's amplitude (equivalently, its energy) decreases at a rate $\vert h_{x0} \vert \sim t^{- 0.58}$, i.e., much steeper than what the cosmological redshift alone would imply, $\vert h_{x0} \vert \sim (A^2 W)^{-1/3} \sim t^{- 0.42}$. This is a very important feature of $ \vert h_{x0} \vert $, suggesting that, apart from Universe expansion, there is an additional descending factor of the GW amplitude, most probably due to the gravitational-energy loss in the interaction between gravitational, EM, and the cosmic-fluid degrees of freedom (they constitute a closed system). Notice that the same is true, also as regards the evolution of $h_{x \xi}$ and $h_{x v}$. 

\begin{figure}[!ht]
\epsscale{0.60} \plotone{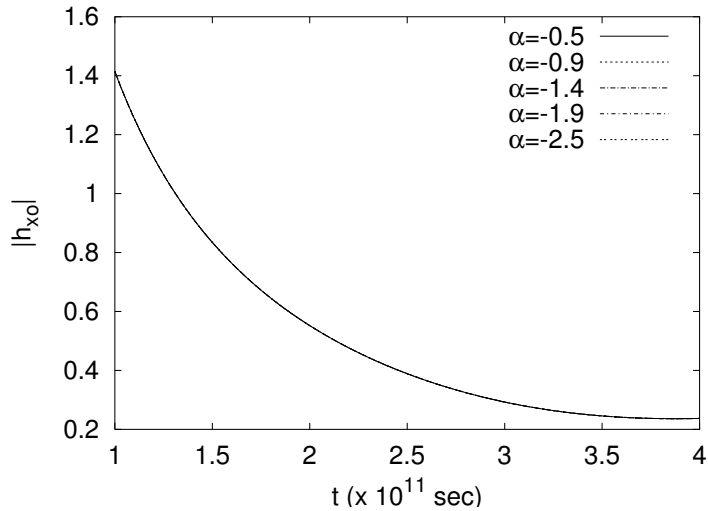} \caption{The time evolution of the normalized (over its initial value) GW amplitude $\vert h_{x0} \vert$, for $\sg = 10 \: sec^{-1}, \; \gm = 0.6, \; \om_g = 0.1 \: Hz$, and $\theta = 45^{\circ}$. The apparent independence of $\vert h_{x0} (t) \vert $ on $\al$, which is used to parametrize its initial value, is, in fact, a consequence of constraint (31). We observe that the metric perturbation's amplitude (equivalently, its energy) decreases at a higher rate than what the cosmological redshift alone would imply, indicating that there is an additional GW-energy loss.}
\end{figure}

A potential energy transfer from gravitational to the EM (and the fluid) degrees of freedom due to the resonant interaction between GWs and MHD waves in curved spacetime is an {\em irreversible process}, and not only because of the resistivity of the cosmic fluid involved (in connection see, e.g., Tsagas 2011, Mavrogiannis and Tsagas 2021). In this case, the subsequent damping of the GW has a significant sideffect, the {\em production of entropy}. The entropy, $ S$, associated to a GW is a natural extension of the quantum {\em von Neumann entropy} into classical, four-dimensional wave systems (see, e.g., Kawamori 2021), i.e., \be S = - \int_{t_0}^t dt \int_V d^3 x \sqrt{-g} \: \vert  h_{x0} \vert^2 \ln \vert h_{x0} \vert^2 \: , \ee where $V$ is the spatial volume involved and $\sqrt{-g} = \left ( \frac{t}{t_0} \right )^{\ell + 1}$ is the determinant of the metric tensor (1). In our case, the GW's profile does not have any spatial dependence, i.e., its amplitude depends only on time. Therefore, the associated {\em entropy density}, ${\cal S} = S / V$, is given by \be {\cal S} = - \int_{t_0}^t \sqrt{-g} \: \vert  h_{x0} \vert^2 \ln \vert h_{x0} \vert^2 dt \: . \ee At the {\em initial state} $(i)$, i.e., in the absence of any interaction between GWs and MHD waves, the amplitude of the GW can be written in the form \be \vert h_{x0}^{(i)} \vert = A \times \left ( \frac{t_0}{t} \right )^{0.42} \: , \ee where $A$ is given by Eq. (55). On the other hand, at the {\em final state} $(f)$ of the irreversible process under consideration, i.e., in the resonant interaction between GWs and MHD waves, we have \be \vert h_{x0}^{(f)} \vert = A \times \left ( \frac{t_0}{t} \right )^{0.58} \: . \ee With the aid of Eqs. (57) - (59), we find that the {\em entropy density variation} between the initial and the final state of the GW evolution in its resonant interaction with MHD waves is given by \bea \Dl {\cal S} = {\cal S}_{f} - {\cal S}_{i} & = & \frac{1}{\ell + 0.84} \: A^2 \left ( \frac{t}{t_0} \right )^{\ell + 0.84} \\ & \times & \left [ \left ( \frac{t}{t_0} \right )^{0.32} \ln \left ( \frac{t}{t_0} \right ) - \frac{\ell + 0.84}{\ell +1.16} \ln \left ( \frac{A}{t_0} \right ) + \left ( 2 \ell + 103.5 \right ) + {\cal O} \left ( A^{0.75} \right ) \right ] \: , \nn  \eea where ${\cal O} \left ( A^{0.75} \right )$ denotes terms of the order $A^{0.75}$ $\left ( \sim 10^{-8} \right )$. For $\gm = 0.6$ $(\ell = 0.25)$ and $1 \leq \frac{t}{t_0} \leq 100$, Eq. (60) results in $ \Dl {\cal S} \: \simeq \: 1.3 \times 10^4 \; A^2 $, which is a definitely positive, although relatively small $\left ( \Dl {\cal S} \sim 10^{-18} \right )$ quantity.

On the other hand, the energy, ${\cal E}_g$, carried by a GW, can not be defined locally. It can only be {\em quasi-localized} under very certain conditions (for more details see, e.g., Cai et al. 2022). Provided that these conditions are met, the energy carried by a GW is, essentially, proportional to the amplitude square of the GW, i.e., ${\cal E}_g \sim \vert h_{x0} \vert^2$. Hence, an estimate of the gravitational energy lost in the resonant interaction between GWs and MHD waves can be given by comparing the amplitude square of the GW at the aforementioned initial $(i)$ and final $(f)$ states of the interaction process. Accordingly, \be \frac{{\cal E}_{g}^{(i)}}{{\cal E}_{g}^{(f)}} \sim \frac{\vert h_{x0}^{(i)} \vert^2}{\vert h_{x0}^{(f)} \vert^2} = \left ( \frac{t}{t_0} \right )^{0.32} = 100^{0.32} = 4.365 \: . \ee In view of Eq. (61), the GW energy at the beginning of the resonant interaction process was more than four times larger than that at the end of this process. To the best of our knowledge, with the exception of large concentrations of plasma around compact objects, there is no direct conversion of gravitational energy into {\em heat}. Therefore, we expect that this energy deficit corresponds to the energy transferred to the EM and the fluid degrees of freedom. But, Eq. (61) reveals also a much more interesting feature, namely, \be {\cal E}_{g}^{(i)} \approx 4 {\cal E}_{g}^{(f)} \Rightarrow \vert h_{x0}^{(i)} \vert \approx 2 \vert h_{x0}^{(f)} \vert \: . \ee In other words, if resonant interaction between cosmological GWs (CGWs) and MHD waves has ever taken place in the expansion history of the Universe, then the observed CGW amplitude at the present epoch would be half than what is otherwise being expected. Such a result would be an indirect, though quite clear manifestation that large scale magnetic fields have indeed played some role in the early Universe evolution.

In response to the aforementioned results, we consider the temporal evolution of the electric-field perturbations $\dl E^z$ (the Alfv\'{e}n mode) and $\dl E^y$ (the magnetosonic mode), parame-trized by the normalized initial amplitude of the GW, $\al$ (Figs. 2). We observe that, as $\al$ increases in absolute value in the range $0.5 \leq \al \leq 2.5$, $\vert \dl E^z \vert$ almost doubles its maximum value, rising as $\vert \dl E^z \vert \sim \al^{0.44}$, while the magnetosonic component remains unaffected by the variation of $\al$, increasing very rapidly to reach at an amplitude 2500 times larger than that of $\dl E^z$. On the other hand, the electric-field perturbation along the $\hat{x}-$axis remains {\em null} $\left ( \vert \dl E^x \vert \simeq 0 \right )$, i.e., it is not amplified effectively during the whole temporal integration interval, $1 \leq \frac{t}{t_0} \leq 100$.

\begin{figure}[!ht]
\epsscale{1.20} \plottwo{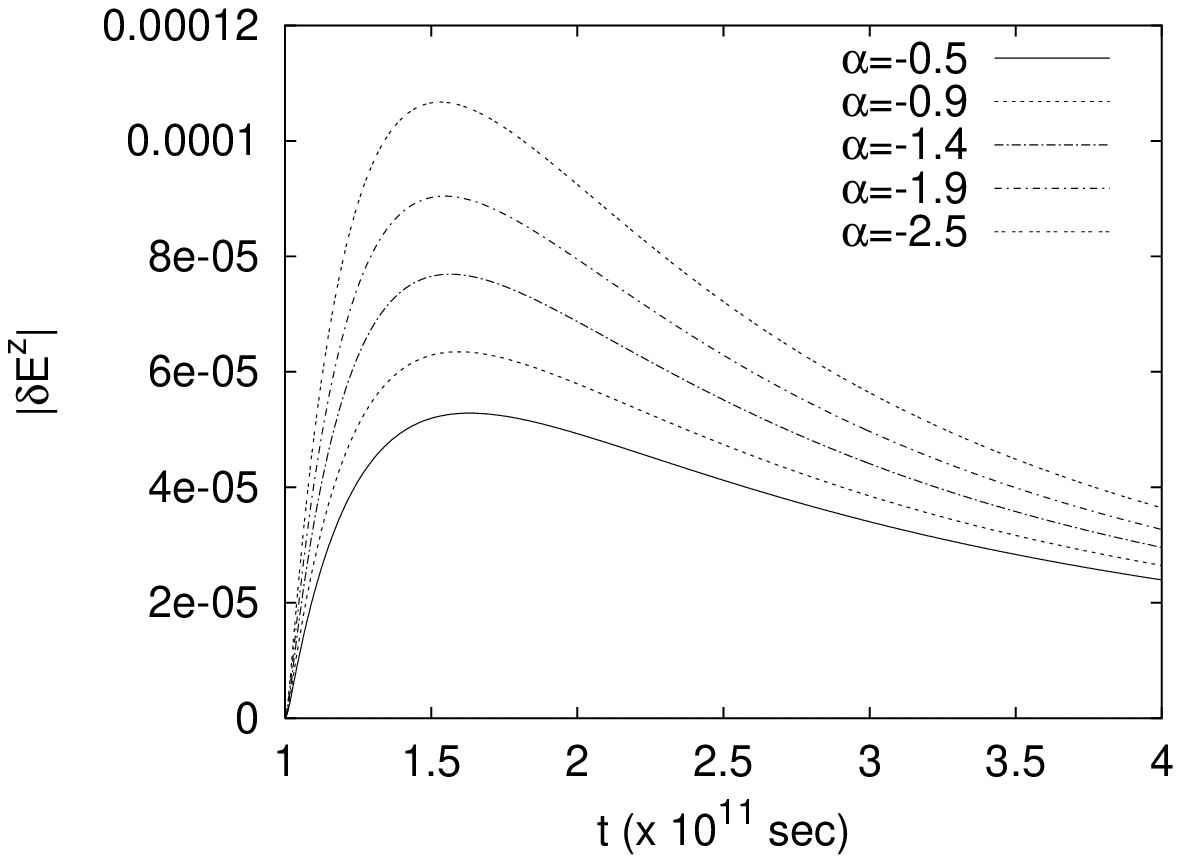}{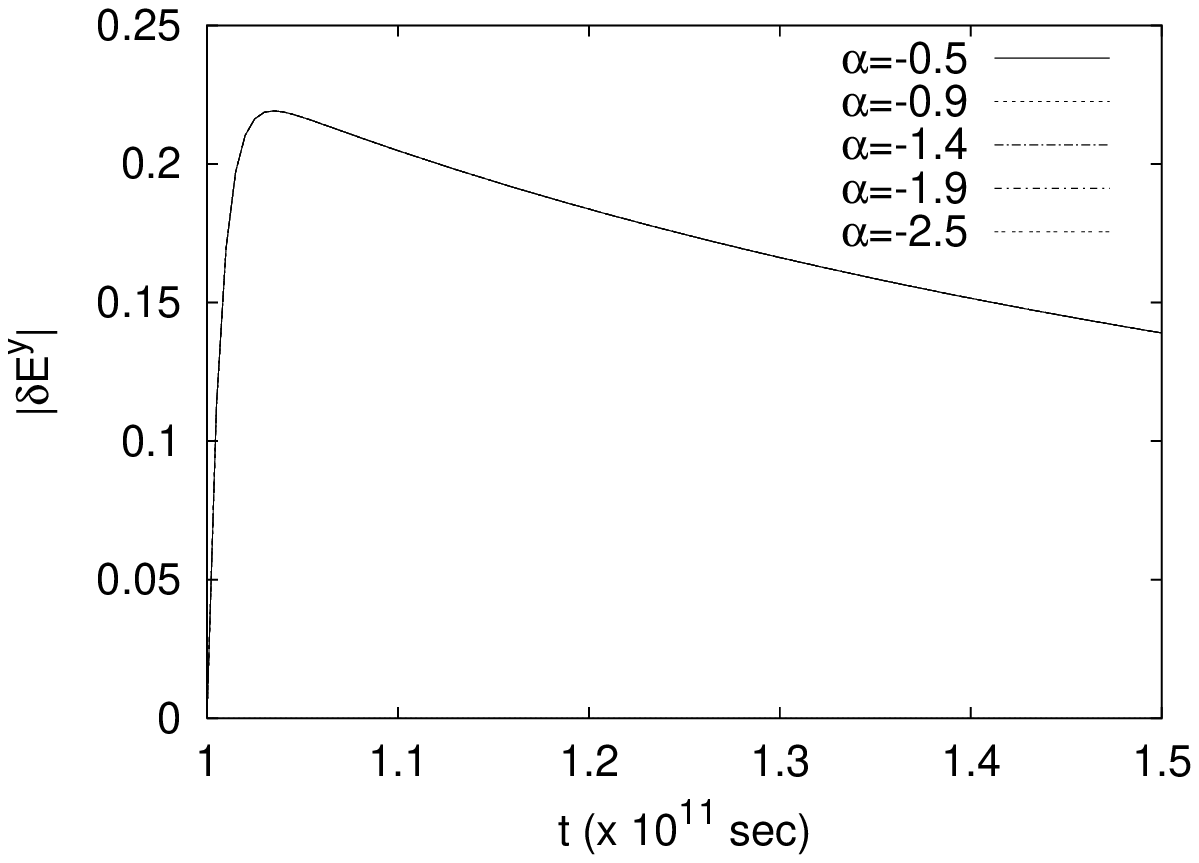} \caption{The electric field perturbations $\dl E^z$ and $\dl E^y$ for $\gm = 0.6, \; \sg = 10.0 \; sec^{-1}, \; \om_g = 0.1 \: Hz$, and $\theta = 45^{\circ}$. We observe that, as regards the evolution of the magnetosonic component, there is no dependence on the normalized intial amplitude of the GW, $\al$, whereas the associated Alfv\'{e}n component rises more steeply as $\al$ increases in absolute value.}
\end{figure}

Now, that we have a clear-cut case regarding GW's propagation in the magnetized plasma that drives the evolution of Thorne's model - i.e., it can certainly transfer a part of its energy to the EM-field and/or the cosmic-fluid degrees of freedom - we may explore the associated response of the various MHD quantities involved. We need to point out that the three linearly independent constraints $c_i = 0$ $(i = 1, 2, 3)$ arising from Eqs. (31), (36), and (43) are also monitored throughout numerical integration. During the whole time interval $1 \leq \frac{t}{t_0} \leq 100$, none of them ever exceeds the value $10^{-10}$, i.e., they remain sufficiently close to zero. This feature suggests that our model is (indeed) self-consistent, thus making us quite confident for the accuracy of our results in spite of the complexity of the associated field equations.

\subsection{Resistive instabilities} 

The first set of parameters to be treated as constants are $\{ \gm = 0.6, \; \om_g = 0.1 \; Hz, \; \theta = 45^{\circ} \}$. In other words, the angle of propagation with respect to the direction of the magnetic field in Thorne's spacetime of fixed anisotropy, as regards a transient, non-dispersive, plane GW of $\al = 1.4$, is also kept fixed at $45^{\circ}$. In this case, the evolution of the various MHD modes - basically the electric-field perturbations $\dl E^z$ (the Alfv\'{e}n component) and $\dl E^y$ (the magnetosonic one) - is parametrized only by the conductivity of the plasma fluid. On the basis of its potential temporal dependence during $t_0 = t_{eq} \leq t \leq t_{rec} = 100 \: t_0$, a representative set of conductivity values would be $\sg = \{1.0, \; 5.0, \; 10.0, \; 15.0, \; 20.0\}$. The corresponding results are shown in Figs. 3. We observe that the electric-field perturbations along the direction of the background magnetic field are, in fact, {\em resistive instabilities}, as they are particularly favoured by small values of conductivity (i.e., high values of resistivity). On the contrary, the perturbations along the (normal) $\hat{y}$-direction (the associated MSWs) increase much more steeply and they are saturated at high amplitude values (almost 2500 times higher than those of the Alfv\'{e}n component) for long enough time intervals $( \Dl t \simeq 10^{12} \: sec)$, in accordance to the increasing conductivity. This is not an unexpected result, since, in the large-conductivity (i.e., low resistivity) limit of an ideal plasma, only the magnetosonic component survives (see, e.g., Kuiroukidis et al. 2007). 

\begin{figure}[!ht]
\epsscale{1.20} \plottwo{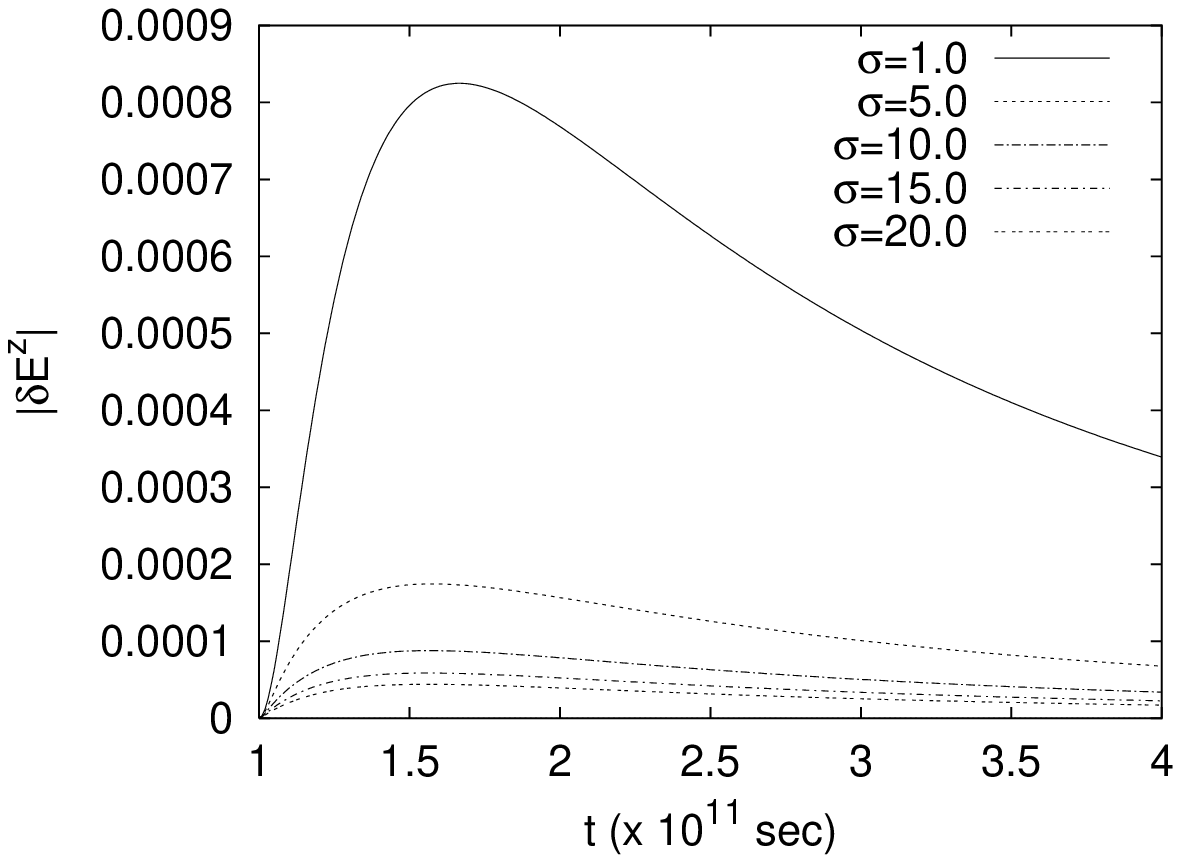}{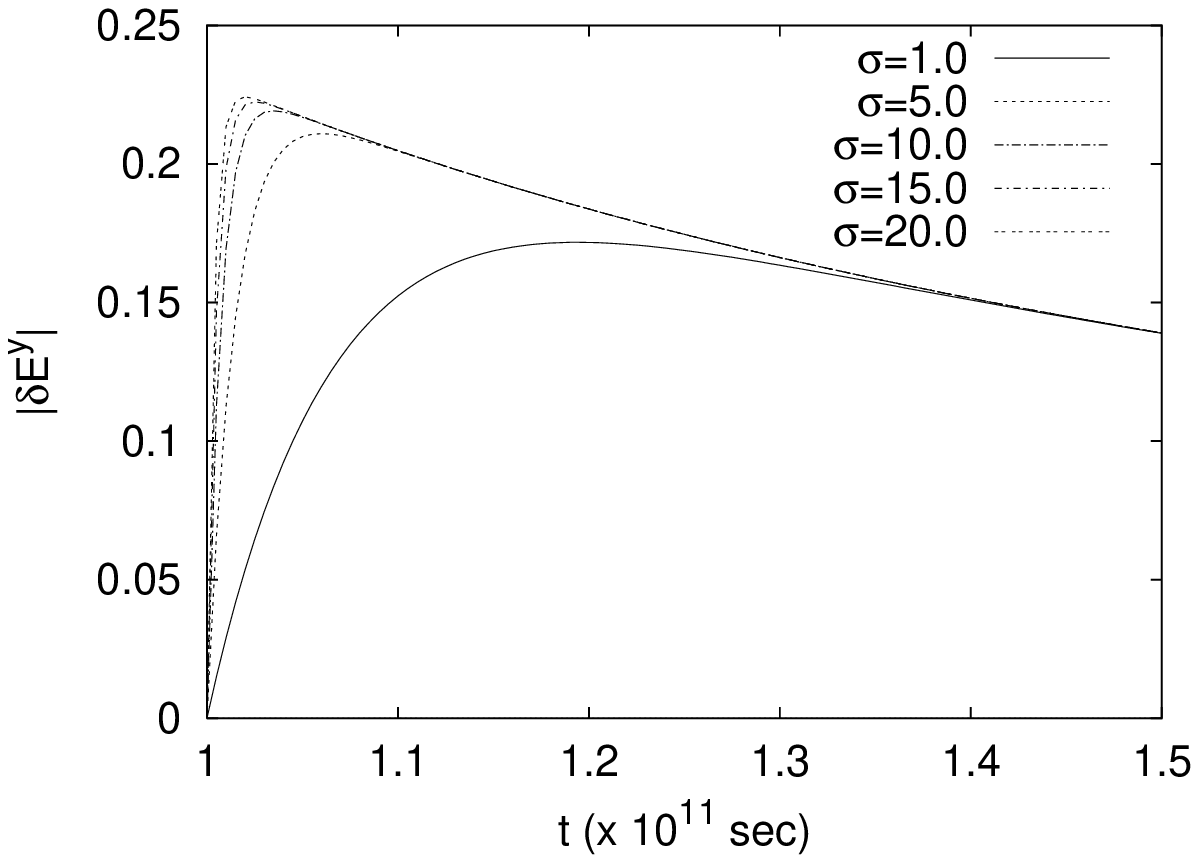} \caption{The temporal evolution of the electric-field perturbations $\dl E^z$ (the Alfv\'{e}n component) and $\dl E^y$ (the magnetosonic component) parametrized by the conductivity, $\sg$, of the plasma fluid, for $\gm = 0.6, \; \om_g = 0.1 \: Hz$, and $\theta = 45^{\circ}$.}
\end{figure}

\vspace{.5cm}

In Fig. 4, the temporal evolution of the desnity perturbations in Thorne's model are presented, as a function of conductivity. It becomes evident that high values of conductivity suppress density fluctuations quite rapidly. In other words, low conductivity values - that clearly signal a departure from the ideal plasma case - do favour mass-density instabilities, a result that is in accordance to earlier works (Kuiroukidis et al. 2007, Kleidis et al. 2008). 

\begin{figure}[!ht]
\epsscale{0.60} \plotone{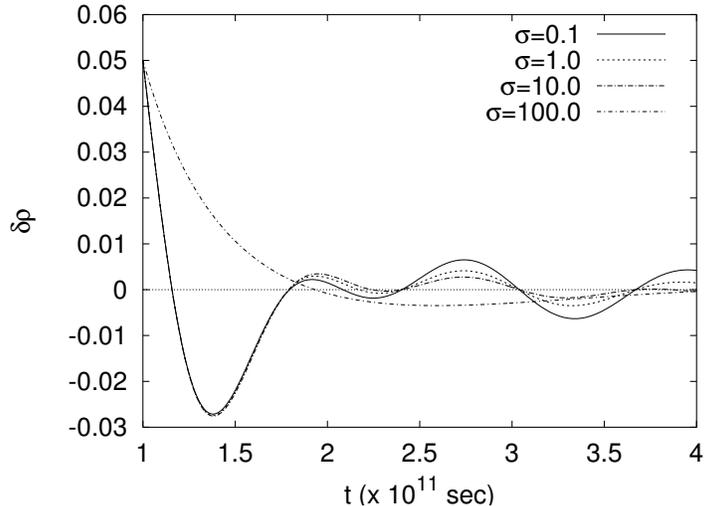} \caption{The temporal evolution of density perturbations for $\gm = 0.6, \; \om_g = 0.1 \: Hz$, and $\theta = 45^{\circ}$, being parametrized by $\sg$. The associated numerical results suggest that large conductivity values suppress mass-density perturbations in a more abrupt manner, i.e., no density instabilities are favoured in an ideal plasma.}
\end{figure}

\subsection{Anisotropic instabilities}

Next, we consider the following set of parameters to be kept constant: $\{ \sg = 10.0 \; sec^{-1}, \; \\ \om_g = 0.1 \: Hz, \; \theta = 45^{\circ} \}$. In this case, the angle of propagation with respect to the direction of the magnetic field in Thorne's spacetime of variable anisotropy (i.e., magnetic-field intesity), as regards a transient, non-dispersive, plane GW, is kept fixed at $45^{\circ}$. The non-ideal plasma that drives the evolution of such a spacetime model has a conductivity of $\sg = 10.0 \: sec^{-1}$. Accordingly, the evolution of the various MHD modes - basically the electric-field perturbations $\dl E^z$ and $\dl E^y$ - is parametrized by the anisotropy measure, $\gm$, taking values of growing anisotropy, i.e., $\gm = \{0.4, \; 0.5, \; 0.6, \; 0.7, \; 0.8\}$. The associated results are presented in Fig. 5.

Once again, the magnetosonic component of the electric-field perturbation increases more steeply than the corresponding Alfv\'{e}n one, reaching at an amplitude almost 3000 times higher than that of $\dl E^z$. In this case, large values of $\gm$ correspond to a higher degree of anisotropy of the background spacetime and to a lower expansion rate along the direction of the ambient magnetic field. For $\gm \geq 0.66$, the magnetic field strength starts to decrease [cf. Eq. (3)] and this behaviour is monitored also in the evolution of the electric field (intersecting curves). Indeed, numerics confirm that, up to $\gm \simeq 0.7$, $\vert \dl E^z \vert$ rises on the increase of the anisotropy measure, whereas, for greater values of $\gm$, this behaviour is reversed (see, e.g., Fig. 6).

\begin{figure}[!ht]
\epsscale{1.20} \plottwo{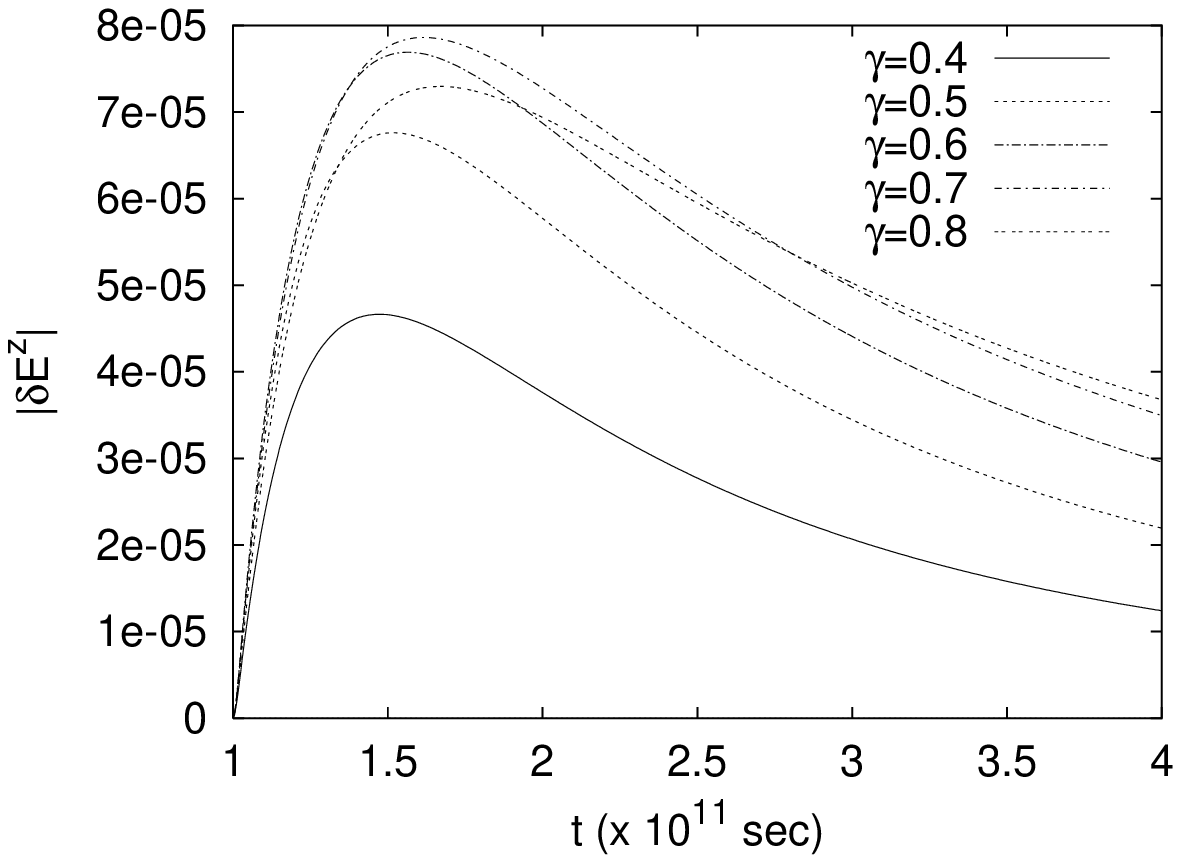}{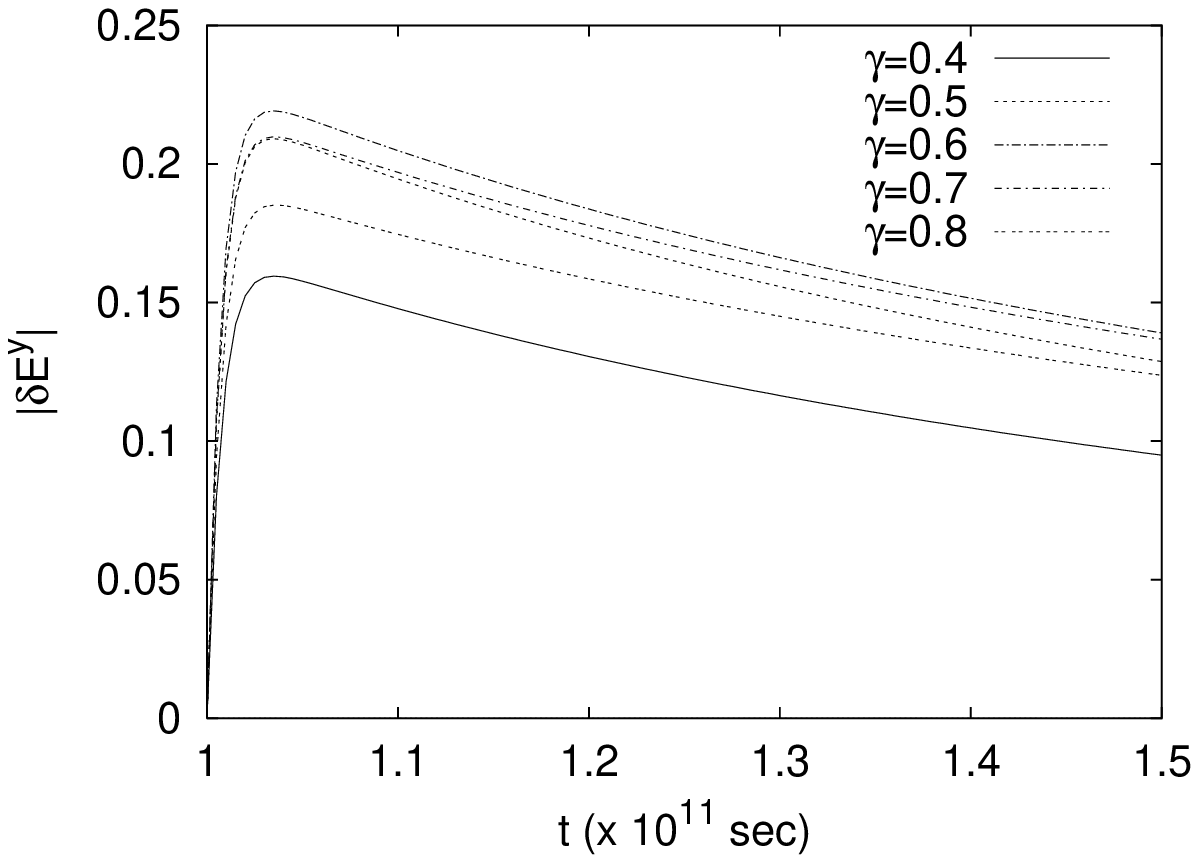} \caption{The electric-field perturbations $\dl E^z$ and $\dl E^y$ for the following set of the parameters involved $\{ \sg = 10.0 \: sec^{-1}, \; \om_g = 0.1 \: Hz, \; \theta = 45^{\circ} \}$, illustrating their dependence on the anisotropy measure, $\gm$.}
\end{figure}
 
\begin{figure}[!ht]
\epsscale{0.60} \plotone{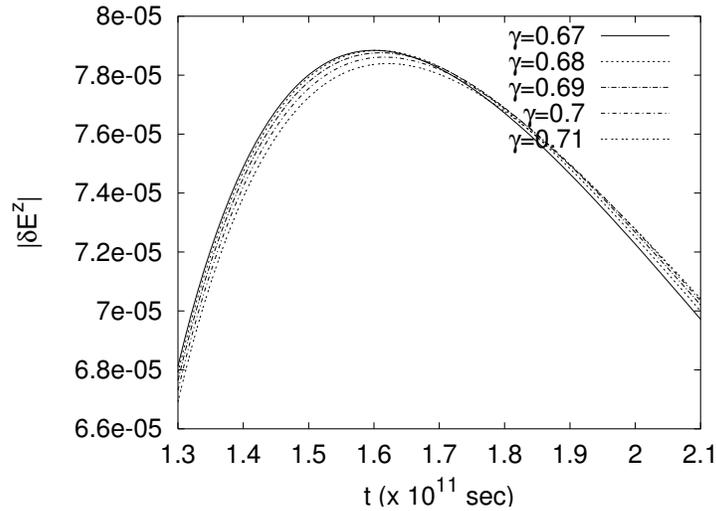} \caption{The electric-field perturbation along the $\hat{z}$-direction for the set of parameters above, verifying numerically that the turning point in the behaviour of the magnetic field is attained at $\gm = 0.67$.}
\end{figure}

In Fig. 7, we illustrate the dependence of the mass-density perturbations on $\gm$. The corresponding results are displayed in a limited time-interval for reasons of best analysis. We observe that, the density perturbations decay more rapidly as $\gm$ grows. Large values of $\gm$ correspond to a slow expansion rate along the direction of the magnetic field. Therefore, we expect that, normal to this direction, condensations that may be formed within the plasma fluid will {\em remain active} for longer time-intervals, something that could lead to {\em pancake instabilities}. It appears that one- and/or two-dimensional formations are actually quite common in magnetized anisotropic cosmological models (see, e.g., Kuiroukidis et al. 2007).

\begin{figure}[!ht]
\epsscale{0.60} \plotone{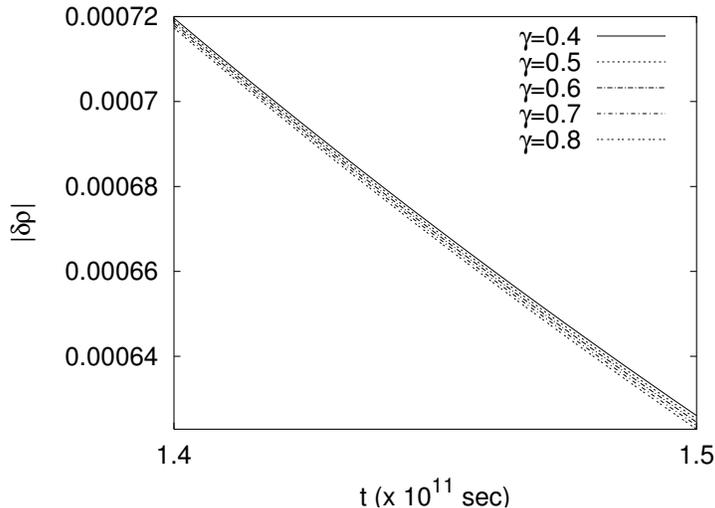} \caption{Density perturbation plots for $\sg = 10.0 \: sec^{-1}, \; \om_g = 0.1 \: Hz$, and $\theta = 45^{\circ}$. Notice that, for large values of $\gm$, the density perturbations decay more prominently.}
\end{figure}

\subsection{Dispersive instabilities} 

Now, we examine the dependence of the temporal evolution of the electric-field perturbations $\dl E^z$ and $\dl E^y$ on the frequency of the GW. To do so, we consider the following set of constants $\{ \gm = 0.6, \; \sg = 10.0 \; sec^{-1}, \; \theta = 45^{\circ} \}$, while $\om_g$ admits the values $\om_g = \{0.01, \; 0.05, \; 0.10, \; 0.15, \; 0.20\}$, measured in $Hz$. This dependence is illustrated in Figs. 8. We observe that the electric-field perturbation along the direction of the background magnetic field is affected by the variation of the frequency, while those along the $\hat{y}$-direction remain unaltered. In particular, the first of Figs. 8 suggests that high-frequency GWs favour an Alfv\'{e}n component that grows linearly with $\om_g$, i.e., $\vert \dl E^z \vert \sim \om_g$. On the other hand, the independence of $\vert \dl E^y \vert$ on $\om_g$ is rather surprising, since, in the isotropic model one would have an $\om_g^2$-dependence (see, e.g., Moortgat and Kuijpers 2003). Instead, here, $\vert \dl E^y \vert$ rises very rapidly to reach values 5000 times higher than the Alfv\'{e}n component and remains almost saturated for a long time interval. This is a typical behaviour of the anisotropic magnetosonic instability that we saw earlier (cf. Fig. 5, for $\gm = 0.6$). It appears that, as regards MSW propagation in the magnetized plasma under consideration, the spacetime anisotropy prevails over any frequency modulation of the GW.

\begin{figure}[!ht]
\epsscale{1.20} \plottwo{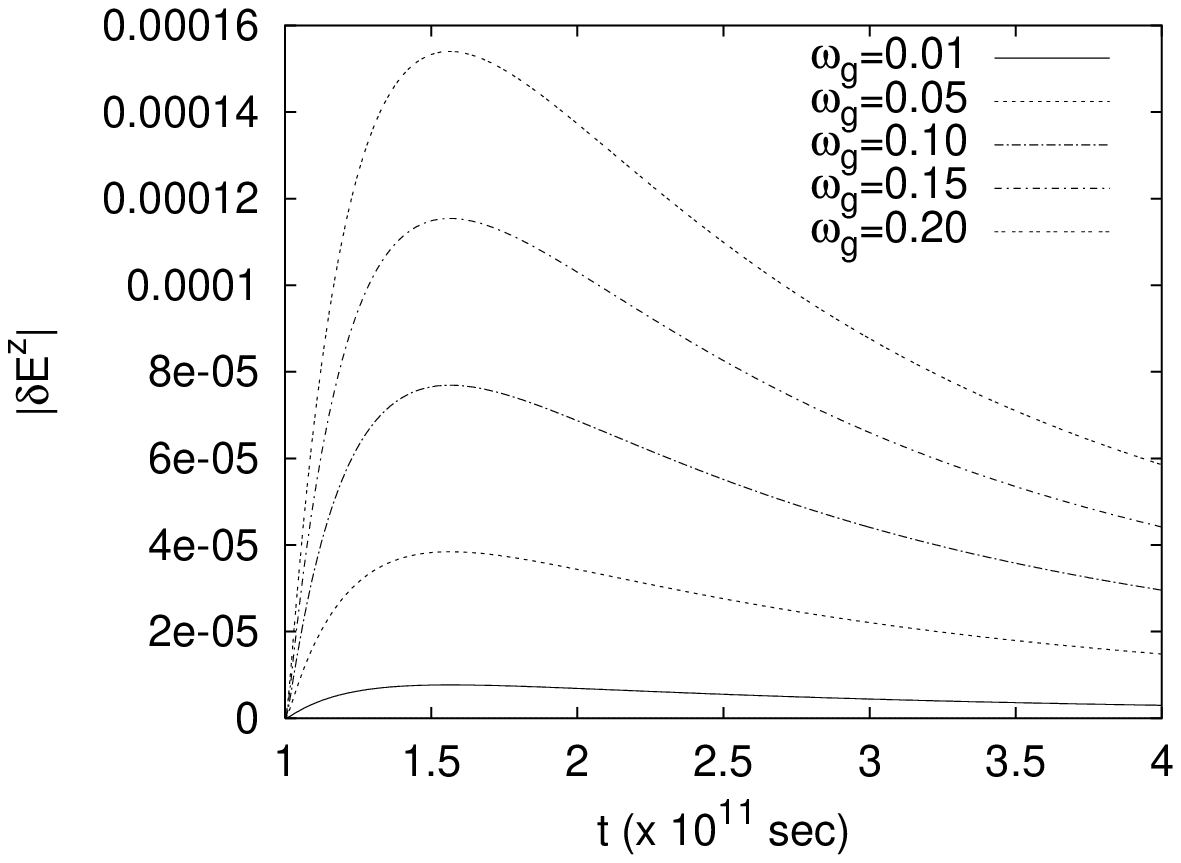}{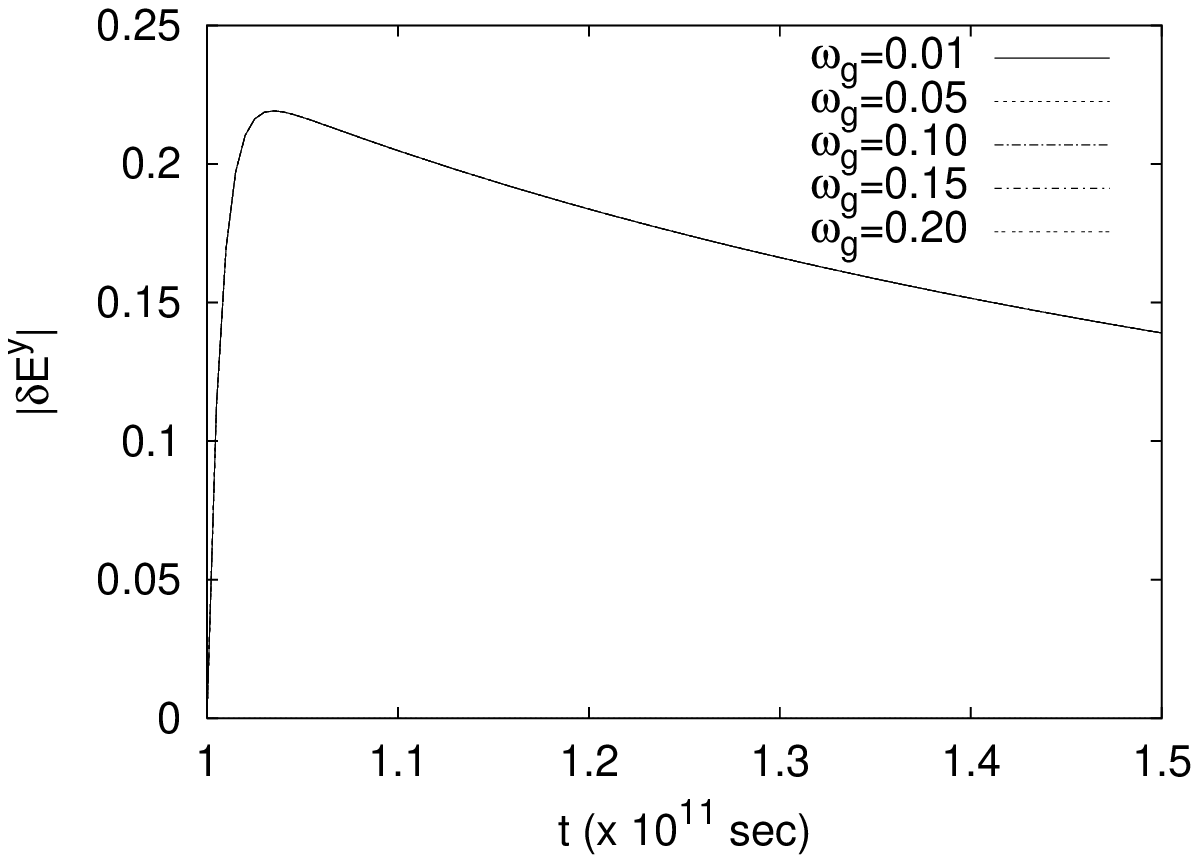} \caption{The temporal evolution of the electric-field perturbations $\dl E^z$ and $\dl E^y$ for the following set of the parameters involved $\{ \gm = 0.6, \; \sg = 10.0 \; sec^{-1}, \; \theta = 45^{\circ} \}$. Although the Alfv\'{e}n component grows as $\dl E^z \sim \om_g$, the magnetosonic one remains unaffected by the variation of $\om_g$.}
\end{figure}

Furthermore, we observe that a decreasing $\om_g$ (i.e., the long GW wavelength regime) also favours condensations, that can be formed within the plasma fluid, to remain active for longer time-intervals (Fig. 9).

\begin{figure}[!ht]
\epsscale{0.60} \plotone{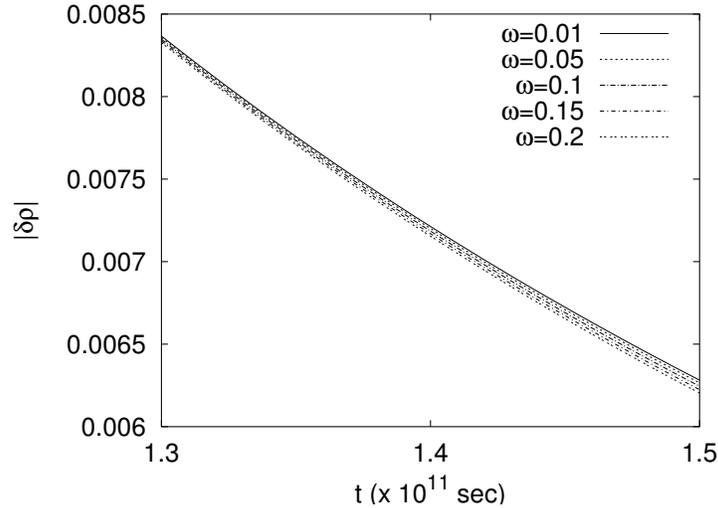} \caption{Density perturbation plots, for $\gm = 0.6, \; \sg = 10.0 \; sec^{-1}$, and $\theta = 45^{\circ}$. Notice that, the higher the frequency of the wave, the faster they decay in time.}
\end{figure}

\subsection{Resonant instabilities} 

To conclude our analysis on the electric-field perturbations, we now consider their evolution with respect to the angle of the GW propagation with respect to the direction of the ambient magnetic field of Thorne's model, i.e., $\theta =\{15^{\circ}, \; 30^{\circ}, \; 45^{\circ}, \; 60^{\circ}, \; 80^{\circ}\}$ (see Figs. 10, where the following set of parameters have been treated as constants: $\{\gm = 0.6, \; \sg = 10.0 \; sec^{-1}, \; \om = 0.1 \: Hz \}$). In this case, in addition to the magnetosonic component $\dl E^y$, which is always present as a result of the anisotropic instability (independently of $\theta$), as we approach to normal propagation ($\theta \rarrow 0^{\circ}$) the Alfv\'{e}n component $(\dl E^z)$ is also excited significantly. The reason is that, when the wave-vector becomes vertical to the direction of the magnetic field, $\dl E^z$ also corresponds to a magnetosonic mode of the system under study.

\begin{figure}[!ht]
\epsscale{1.20} \plottwo{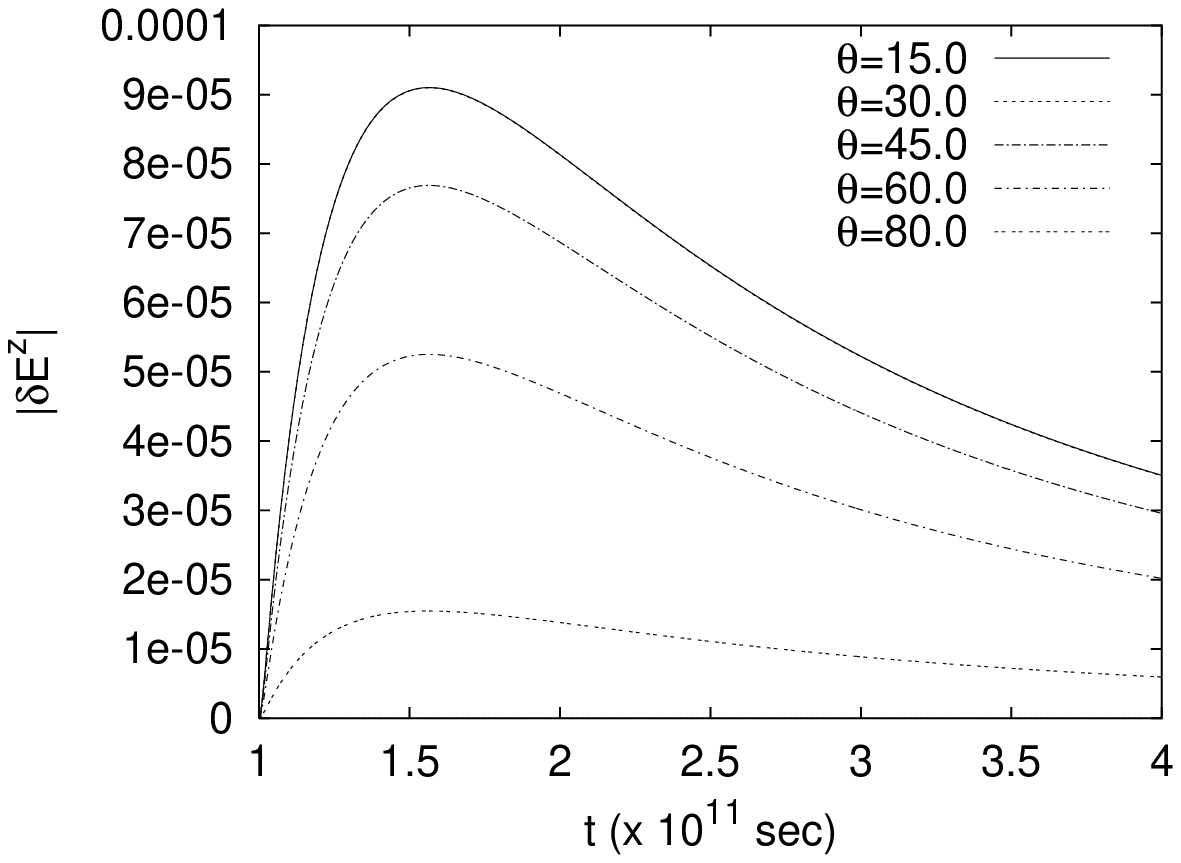}{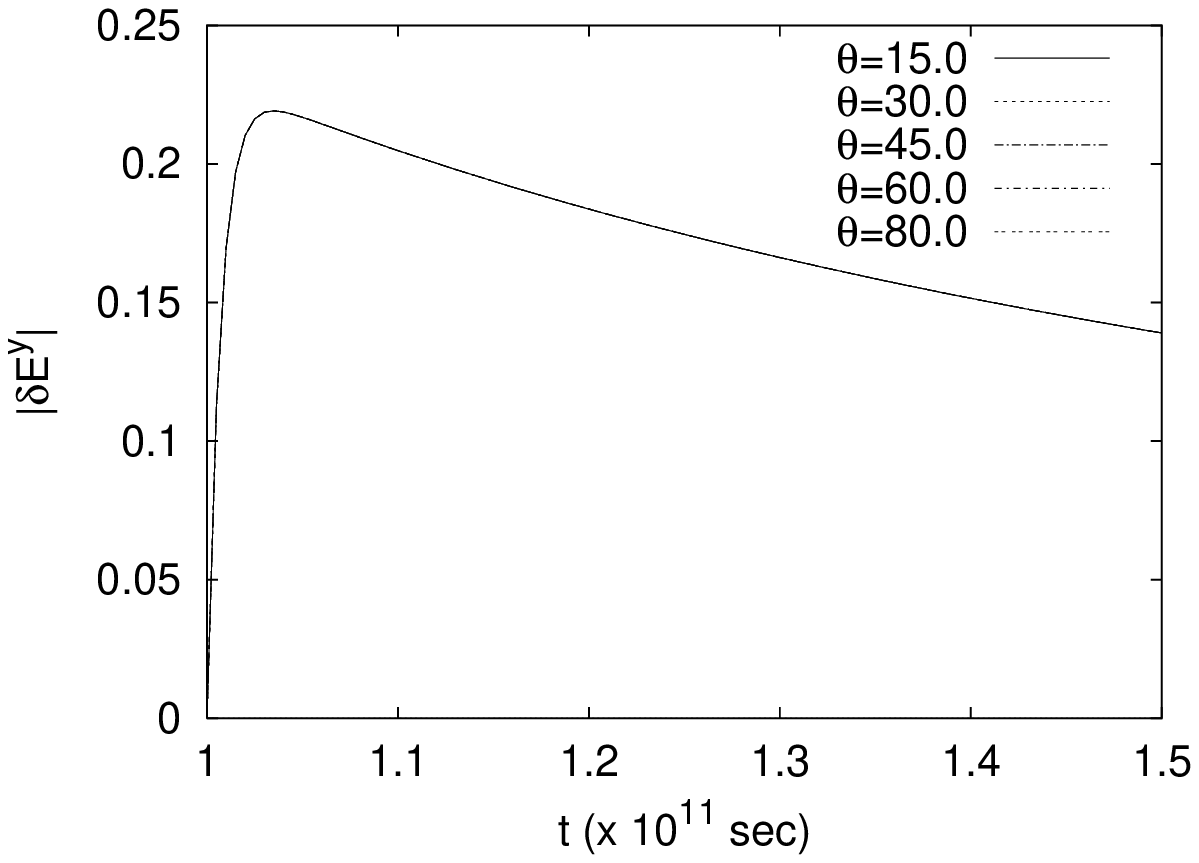} \caption{The temporal evolution of the electric-field perturbations $\dl E^z$ and $\dl E^y$ for the following set of the parameters involved $\{ \gm = 0.6, \; \sg = 10.0 \; sec^{-1}, \; \om_g = 0.1 \: Hz \}$, versus the angle of GW propagation, $\theta$.}
\end{figure}

\subsection{Magnetic-field perturbations} 

Finally, in Figs. 11 and 12 we depict the temporal evolution of the magnetic-field perturbations parametrized by the normalized initial amplitude of the GW, $\al$, the conductivity of the plasma fluid, $\sg$, and the angle of propagation of the metric perturbation, $\theta$. The numerical results indicate that, in this case, only the $\hat{x}$-component of the magnetic-field perturbations (the associated MSW) is excited significantly, probably due to the anisotropic instability that is always present (in connection, see, e.g., Kuiroukidis et al. 2007). Nevertheless, $\dl B^x$ exhibits also some other, very interesting features. In particular, as $\al$ increases in absolute value in the range $0.5 \leq \al \leq 2.5$, $\vert \dl B^x \vert$ almost doubles its maximum value, rising as $\vert \dl B^x \vert \sim \al^{0.44}$. This is exactly the same behaviour as in the case of the Alfv\'{e}n electric-field perturbation (see Figs. 2). It appears that the magnetic-field MSWs do get triggered by the GW, while the corresponding electric-fileld components are driven mainly by the background anisotropy. The opposite behaviour is excibited by the associated Alfv\'{e}n components. Finally, the magnetosonic component $\dl B^x$ appears to be independent of $\sg$, but not of the angle of propagation, $\theta$, with respect to which it grows as $\vert \dl B^x \vert \sim \theta^{\: 0.22}$, i.e., it becomes even more prominent on the approach to the parallel propagation case (cf. Fig. 12).

\begin{figure}[!ht]
\epsscale{1.20} \plottwo{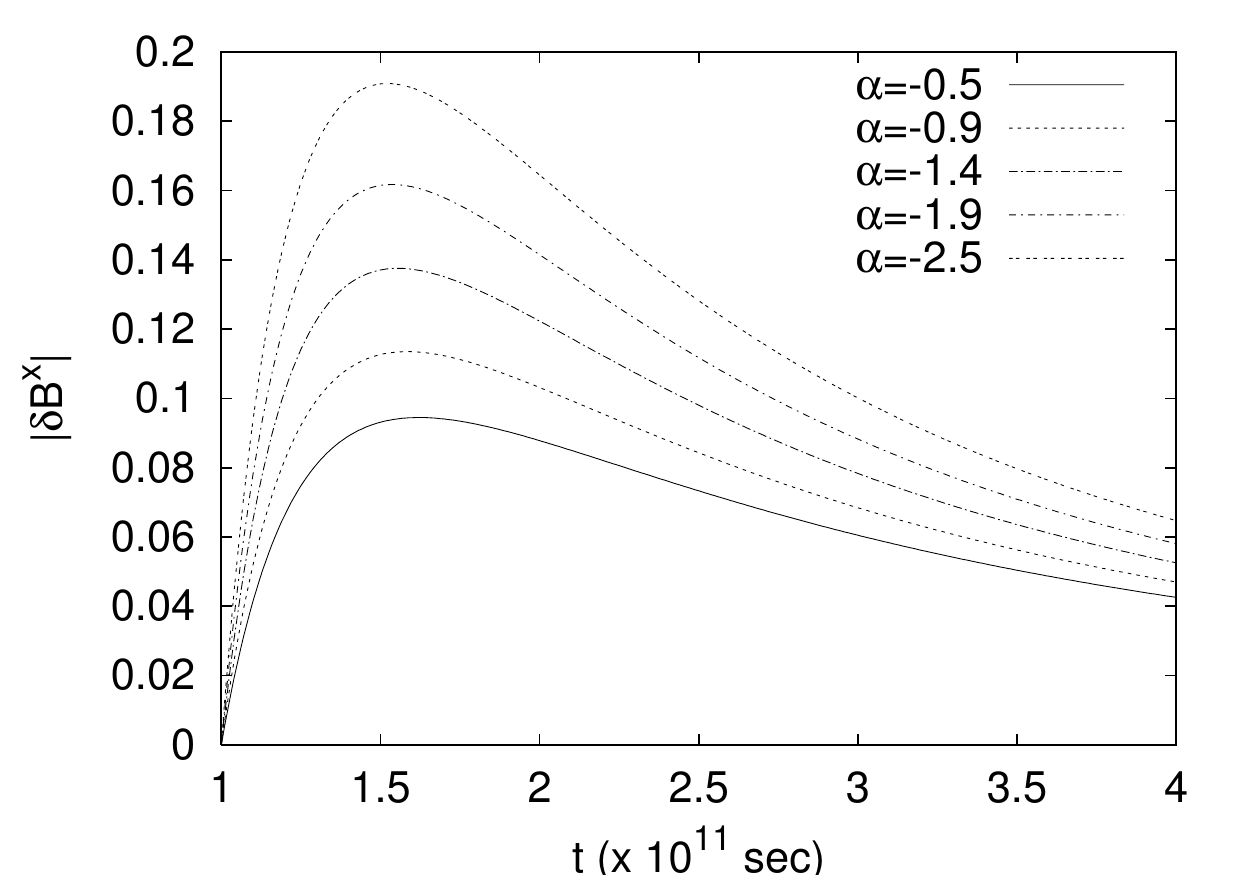}{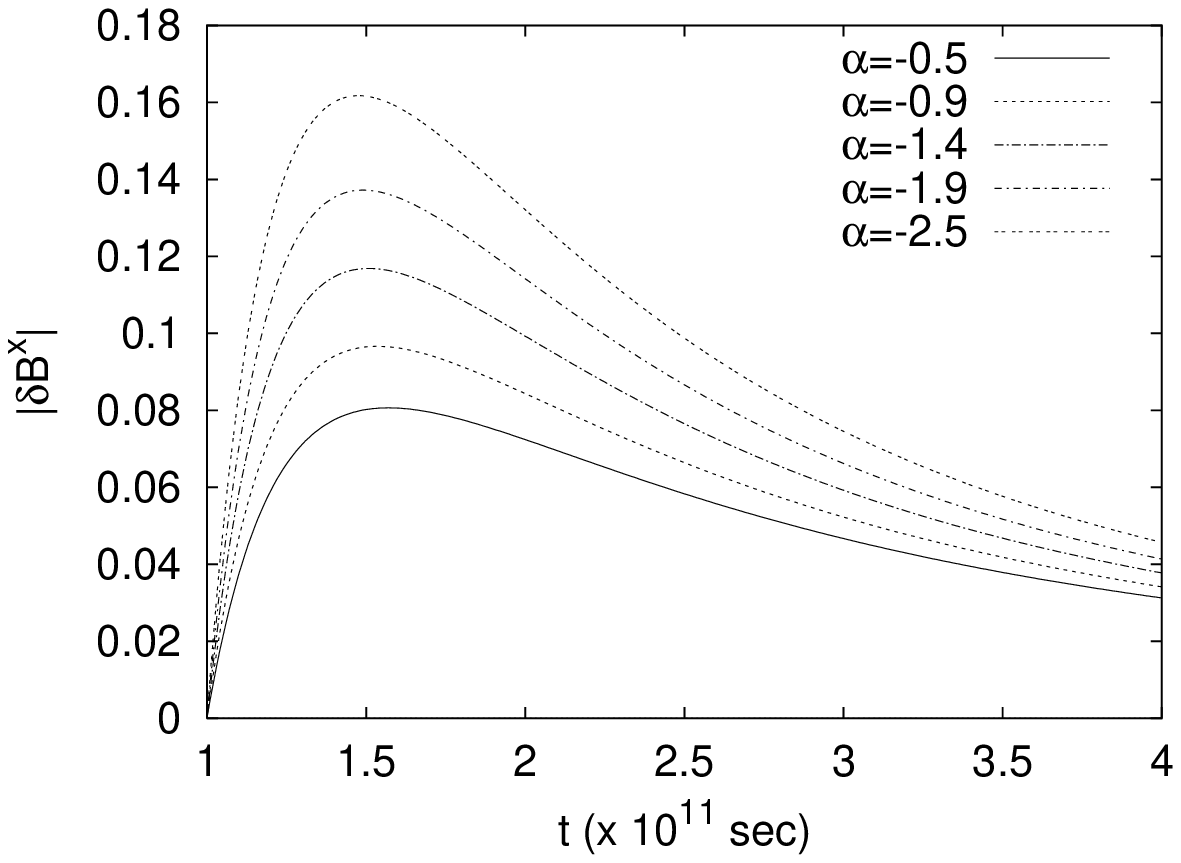} \caption{The magnetic-field perturbation $\dl B^x$: In the first figure we use the set of constant $\{\gm = 0.6, \; \sg = 10.0 \; sec^{-1}, \; \theta = 45^{\circ}, \; \om_g = 0.1 \: Hz \}$, while, the second one refers to $\{ \gm = 0.8, \; \sg = 20.0 \; sec^{-1}, \; \theta = 30^{\circ}, \; \om_g=0.05 \: Hz \}$. We observe that, in both cases the behaviour of the magnetosonic component is almost identical, i.e., $\dl B^x$ is triggered basically by the GW, and in particular, as $\vert \dl B^x \vert \sim \al^{0.44}$.}
\end{figure}

\begin{figure}[!ht]
\epsscale{1.20} \plottwo{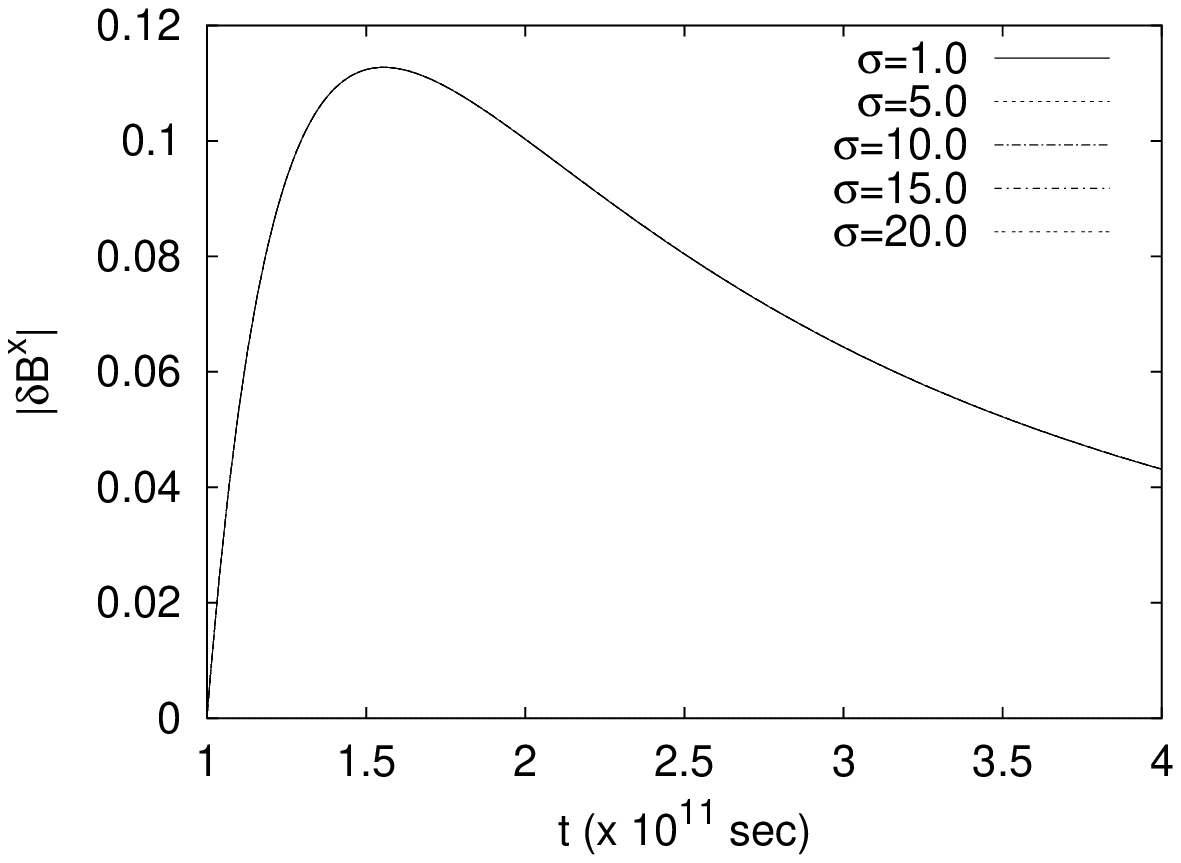}{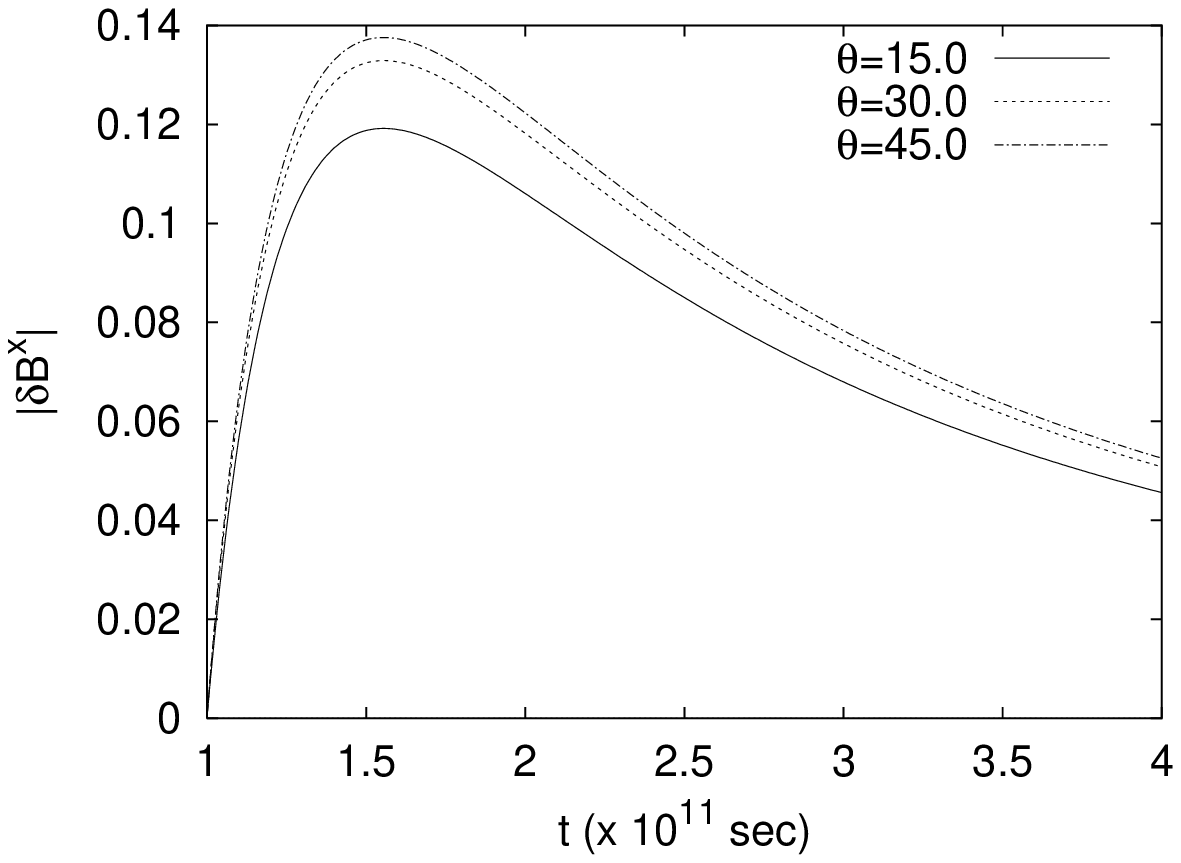} \caption{The temporal evolution of the magnetosonic mode $\dl B^x$ for $\{ \gm = 0.6, \; \al = 1.4, \; \theta = 45^{\circ}, \; \om_g = 0.1 \: Hz \}$ and $\{ \gm = 0.6, \; \sg = 10.0 \; sec^{-1}, \; \al = 1.4, \; \om_g = 0.1 \: Hz \}$. We observe that, it is independent of $\sg$, i.e., in the magnetized plasma fluid that drives the evolution of Thorne's model with $\gm = 0.6$ no resistive magnetic instabilities ever occur. On the contrary, the growth of $\dl B^x$ becomes even more pronounced as $\theta$ grows, i.e., on the approach to parallel propagation with respect to the direction of the ambient magnetic field.}
\end{figure}

\section{Discussion and conclusions}

Using the resistive MHD equations in curved spacetime, we investigated numerically the temporal evolution of the electric- and magnetic-field perturbations, along with the associated mass-density ones, that can be triggered by the oblique propagation of a plane-polarized GW in a class of anisotropic cosmological models with an ambient magnetic field along the $\hat{z}$-axis, namely, in Thorne's cosmological model. This model could be a viable extension to the Standard Model of the Early Universe in the presence of a large-scale magnetic field, since such a field can affect the local spacetime structure; accordingly, an anisotropic background must be taken into account to guarantee the proper treatment of curved spacetime. 

Technically speaking, the GWs affect the evolution of MHD waves in curved spacetime through the general-relativistic Euler equations of motion (18) and the associated Maxwell field equations (19). The {\em gravito-electromagnetic} interaction so assumed is an irreversible process, mostly due to the {\em backreaction} of the EM fields and the fluid on the curved spacetime background, through the Einstein field equations (14) - (16). In this article, by admitting a perfect fluid source, we have actually neglected potential distortions of the curved spacetime background due to energy flows. In general, one could use a {\em non-ideal gas}, where {\em heat conduction, shear} and {\em viscosity} of the fluid source should also be taken into account when solving the Einstein field equations. In fact, it would be particularly interesting to examine what would be the exact nature of the aforementioned gravito-electromagnetic interaction in a realistic (i.e., non-ideal) fluid. This will be the scope of a future work. 

In this article we are mainly interested in the resonant interaction between GWs and MHD waves, i.e., when the value of the MHD circular frequency is close to the corresponding GW quantity. As we have found, in this case energy transfer from the gravitational to the EM degrees of freedom does take place, resulting in the excitation of the latter and in the damping of the GW. The various MHD modes excited, increase quite rapidly at early times after $t_0$ (where the main resonance $\om \simeq \om_g$ takes place), to reach at a maximum value, while, afterwards, they decay at a much lower rate. In some cases, the perturbations (basically, the magnetosonic modes) are {\em saturated} at high amplitude values, for long enough time-intervals, driven by the inherent background anisotropy. In particular, we have identified the following major points:

When a plane-polarized GW propagates in a conductive plasma fluid, oblique-ly to the direction of the ambient magnetic field that drives the evolution of Thorne's model, its amplitude (equivently, its energy) decreases at a rate much higher than what the cosmological redshift alone would imply (Fig. 1). This result suggests that, apart from Universe expansion, there is an additional descending factor of the GW amplitude, most probably due to the gravitational energy lost in the interaction between gravitational, EM, and the cosmic-fluid degrees of freedom, since all of them constitute a closed system. In this context, the GW damping results also in the production of an extra amount of entropy. Indeed, in the resonant interaction between GWs and MHD waves in Thorne's model, the entropy density variation, $\Dl {\cal S}$, between the initial state (when there is no interaction at all) and the corresponding final one (i.e., at the end of the gravito-electromagnetic interaction process) is directly proportional to the amplitude square of the GW involved [cf. Eq. (61)], which is a definitely positive, although quite small quantity.

Certainly, the energy carried by a GW can not be defined locally. However, an estimate of the gravitational energy that is lost in the resonant interaction between GWs and MHD waves can be given, by comparing the amplitude square of the GW at the aforementioned initial and final states of the interaction process. Accordingly, we have found that the GW energy at the beginning of the resonant interaction process was more than four times larger than that at the end of this process [cf. Eq. (62)]. To the best of our knowledge, there is no direct conversion of gravitational energy into heat. Therefore, we expect that this energy deficit corresponds to the energy transferred to the EM and the fluid degrees of freedom. This result, however, might have a much more intriguing consequence, that is, if a resonant interaction between cosmological GWs (CGWs) and MHD waves has ever taken place in the history of Universe expansion, then the observed CGW amplitude at the present epoch would be half than what is expected [cf. Eq. (63)]. Verification of this result by observations would be an indirect, though quite clear manifestation that large scale magnetic fields have indeed played some role in the early Universe evolution.

In view of the aforementioned results, the temporal evolution of the electric-field perturbations $\dl E^z$ (the Alfv\'{e}n mode) and $\dl E^y$ (the magnetosonic mode) has been examined, parametrized by the normalized initial amplitude of the GW, $\al$ (Figs. 2). We have found that, as $\al$ increases in the range $0.5 \leq \al \leq 2.5$, the Alfv\'{e}n mode $\vert \dl E^z \vert$ rises as $\vert \dl E^z \vert \sim \al^{0.44}$. As regards the associated magnetosonic one, $\dl E^y$, it appears to remain unaffected by the variation of $\al$ and increases rapidly after $t_0$, probably because it is trapped in the second resonance, $\om \simeq 1.005 \: \om_g$. Its amplitude reaches at values 2500 times higher than those of $\dl E^z$, where it saturates for a relatively-long time interval, probably due to the inherent spacetime anisotropy (anisotropic instability). 

On the other hand, as regards the magnetic-field perturbations, numerics indicate that only the $\hat{x}$-component (the associated MSW) is excited significantly, probably due to the anisotropic instability that is always present. The amplitude of $\dl B^x$ also rises with the increasing $\al$, as $\vert \dl B^x \vert \sim \al^{0.44}$, i.e., at exactly the same rate as $\vert \dl E^z \vert$. It appears that the magnetic-field MSWs do get triggered by the GW, while the corresponding electric-fileld components are driven mainly by the background anisotropy. Accordingly, we can not help but wondering whether the ever-present mangetic field in the Universe has been partly achieved at the expense of gravitational radiation. Clearly, this could be the scope of a future work. Notice also that, the magnetosonic component $\dl B^x$ appears to be independent of $\sg$, but not of the angle of propagation, $\theta$, with respect to which its amplitude grows as $\vert \dl B^x \vert \sim \theta^{\: 0.22}$, i.e., it becomes even more prominent on the approach to the parallel propagation case (cf. Fig. 12).

As for the response of the electric components to the variation of the cosmological parameters involved, such as the conductivity of the plasma fluid (resistive instabilities), the anisotropy measure of the curved spacetime (anisotropic instabilities), the frequency modulation of the GW (dispersive instabilities) and the variation of the associated angle of propagation with respect to the direction of the magnetic field (resonant instabilities), it can be summarized as follows: 

{\em Resistive instabilities:} Numerical results suggest that the electric-field perturbations along the direction of the background magnetic field are particularly favoured by low conductivity (i.e., high resistivity) values (Figs. 3). As regards the perturbations along the (normal) $\hat{y}$-direction (the associated MSWs), they increase much more steeply and they are saturated at high amplitude values (almost 2500 times higher than those of the Alfv\'{e}n component) for much longer time intervals $\left ( \Dl t \simeq 10^{12} \: sec \right )$, in accordance to growing conductivity. This is not an unexpected result, since, in the large-$\sg$ limit (i.e., on the approach to the ideal plasma case), only the magnetosonic component survives. 

{\em Anisotropic instabilities:} When anisotropy grows, the magnetosonic component of the electric-field perturbation increases steeply and its amplitude reaches at values almost 3000 times higher than those of $\dl E^z$. Large values of the associated measure,$\gm$, correspond to a lower expansion rate along the direction of the ambient magnetic field. For $\gm \geq 0.66$, the magnetic field strength also starts to decrease [cf. Eq. (3)] and this behaviour is monitored also in the evolution of the electric components (cf. Fig. 5). Indeed, the numerical results confirm that, up to $\gm \simeq 0.7$, $\vert \dl E^z \vert$ rises on the increase of the anisotropy measure, whereas, for greater values of $\gm$, this behaviour is reversed (cf. Fig. 6). 

{\em Dispersive instabilities:} As regards the dependence of the electric modes on the frequency of the GW, we have found that high-frequency GWs favour the Alfv\'{e}n component, which grows linearly with $\om_g$. On the contrary, $\dl E^y$ appers to be independent of $\om_g$, rising steeply to reach at values 5000 times higher than those of $\dl E^z$ and remains saturated for a relatively long time interval. This is a typical behaviour of the anisotropic magnetosonic instability; hence, as regards the propagation of MSWs in the magnetized plasma under consideration, the spacetime anisotropy prevails over any frequency modulation of the GW.

{\em Resonant instabilities:} In this case, in addition to the magnetosonic component $\dl E^y$, which is always present (independently of $\theta$), as we approach to normal propagation ($\theta \rarrow 0^{\circ}$) the Alfv\'{e}n component $\dl E^z$ is also excited significantly. The reason is that, when the wave-vector becomes vertical to the direction of the magnetic field, $\dl E^z$ also corresponds to a magnetosonic mode of the system.

{\em Rest-mass density fluctuations:} Numerics indicate that the rest-mass density perturbations decay more rapidly as $\gm$ grows. Large values of $\gm$ correspond to a low-expansion rate along the direction of the magnetic field; hence, we expect that, normal to this direction, condensations that may be formed within the plasma fluid will remain active for longer time-intervals, something that could lead to a two-dimensional (pancake) instability. On the other hand, large conductivity values suppress mass-density fluctuations quite rapidly. In other words, low conductivity - that signal a departure from the ideal plasma case - may favour mass-density instabilities (resistive Jeans instabilities). Finally, long-wavelength GWs (i.e., of low $\om_g$ values) also favour density fluctuations to remain active for long time intervals.

To the best of our knowledge, this is the first time that all these kinds of instability are collectively examined in Thorne's model, a viable extension to the Standard Model of Universe expansion. Clearly, a comprehensive study regarding the excitation of MHD modes (and their subsequent temporal evolution) by GWs in curved spacetime, not only is far from being exhausted, but, in fact, looks very promising. Therefore, the interaction between gravitational and MHD waves in curved spacetime should be further explored and scrutinized in the search for a most accurate profile of many astrophysical and/or cosmological processes.

\acknowledgements

The authors would like to thank Professor Loukas Vlahos, Dr. Heinz Ishliker, and Professor Christos G. Tsagas for several useful discussions. They would also like to thank the two anonymous referees for their comments and their suggestions, that greatly improved the content of the article.

\end{document}